\documentclass[twocolumn,iop,numberedappendix,appendixfloats]{openjournal}
\usepackage[utf8]{inputenc}
\usepackage{amsmath,amssymb,amstext}
\usepackage{graphicx}
\graphicspath{ {./images/} }
\usepackage{multirow}
\usepackage{natbib}
\usepackage{subcaption}

\newcommand{\rvw}[1]{{#1}}
\newcommand{\corr}[1]{{#1}}

\date{\today}

\begin{document}
\journalinfo{The Open Journal of Astrophysics}
\submitted{submitted February 29, 2024; accepted October 14, 2024}

\shorttitle{The GCS of NGC\,3640}
\shortauthors{Ennis, Caso \& Bassino}
\title{Imprints of interaction processes in the globular cluster system of NGC\,3640}
\author{Ana I. Ennis$^{\star1,2}$}
\author{Juan P. Caso$^{\dagger 3, 4, 5}$}
\author{Lilia P. Bassino$^{\ddagger 5}$}

\affiliation{$^{1}$Waterloo Centre for Astrophysics, University of Waterloo, 200 University Ave W, Waterloo, Ontario N2L 3G1, Canada}
\affiliation{$^{2}$Perimeter Institute for Theoretical Physics, Waterloo, Ontario N2L 2Y5, Canada}
\affiliation{$^{3}$Instituto de Astrof\'isica de La Plata (CCT La Plata -- CONICET, UNLP), Paseo del Bosque S/N, B1900FWA La Plata, Argentina}
\affiliation{$^{4}$Consejo Nacional de Investigaciones Cient\'ificas y T\'ecnicas, Godoy Cruz 2290, C1425FQB, Ciudad Aut\'onoma de Buenos Aires, Argentina}
\affiliation{$^{5}$Facultad de Ciencias Astron\'omicas y Geof\'isicas de la Universidad Nacional de La Plata, Paseo del Bosque S/N, B1900FWA La Plata, Argentina}

\begin{abstract}
We present a wide-field study of the globular cluster systems (GCS) of the elliptical galaxy NGC\,3640 and its companion NGC\,3641, based on observations from Gemini Multi-Object Spectrograph/Gemini \rvw{using the $g'r'i'$ filters}. NGC\,3640 is a shell galaxy which presents a complex morphology, which previous studies have indicated as the sign of a recent `dry' merger, although whether its nearest neighbour could have had an influence in these substructures remains an open question. In this work, we trace the spatial distribution of the globular clusters (GCs) as well as their colour distribution, finding a potential bridge of red (metal-rich) GCs that connects NGC\,3640 to its less massive companion, and signs that the blue (metal-poor) GCs were spatially disturbed by the event that created the shells. \rvw{While NGC\,3640 presents a typical bimodal colour distribution, the GCS attributed to NGC\,3641 appears unimodal, with most GCs presenting an intermediate colour in comparison to those closer to its companion.}
\end{abstract}

\maketitle

\section{Introduction} \label{sec:intro}

Mergers are essential parts of the current $\Lambda$ Cold Dark Matter paradigm, in which dark matter haloes and galaxies are mainly formed through hierarchical assembly \citep{Peebles1982,Blumenthal1984,Davis1985}. In particular, the early evolution of early-type galaxies (ETGs) is dominated by major mergers involving large amounts of gas \citep[][]{Naab2007}. Once they become quiescent, they continue to grow their mass and size through minor dry mergers, with little to no star formation \citep[e.g][]{Naab2009}. Aside from being the driving forces behind the mass growth of a galaxy, mergers shape the morphology of galaxies \citep[][e.g]{Hopkins2010,Kannan2015}, feed central supermassive black holes, and potentially trigger central starbursts \citep{Ellison2011,Satyapal2014}. As such, they are important components of any galaxy formation model. 

The frequency of major mergers experienced by a galaxy is influenced by the local density where it resides, as evidenced by the dependence of the fraction of passive galaxies and stellar mass functions with it \citep[e.g.][]{mcnaught14,etherington17}.
Although this relation is nuanced, it is clear that higher-density environments show larger rates of galaxy mergers, with groups being the most active type of environment.

Tidal features are the most direct evidence of recent mergers, and they have been widely used to characterize the impact of these events on the properties of galaxies. Photometric analysis has shown, for example, that blue ETGs are more likely to present morphological disturbances \citep[][e.g]{Tal2009,Kaviraj2011}, hinting at the presence of younger stellar populations as a consequence of recent mergers. However, most of the signatures of accretion events are found in the form of low surface brightness structures, which require long exposures to be analysed. 

\rvw{Globular clusters (GCs) are considered fossil records of the stellar formation of their host galaxy due to their old ages, $10-13$\,Gyr, \citep[e.g.][]{Strader2005,Tonini2013,Fahrion2020a}. They are also tracers of the evolutionary history of the galaxy, since they are survivors of its accretion events and disruption. GCs are almost ubiquitous in galaxies over $10^{9}\,M_{\odot}$ \citep{Harris2015}, and the properties of these globular cluster systems (GCS) have been linked to those of the halo mass \citep{Hudson2014,Forbes2018,El-Badry2019} and to the environment in which the galaxy is located \citep{DeBortoli2022}. These scaling relations are widely accepted and further the use of GCS as tracers of the merger histories of galaxies in the Local Universe \citep[e.g][]{Sesto2018,Kruijssen2020}. In addition, GCs are intrinsically bright objects, which makes them detectable at distances of up to hundreds of Mpc \citep[e.g][]{AlamoMartinez2013,Harris2024}.} In regions where the stellar halo is too faint, GCs have been used as tracers of stellar streams and tidal tails \citep{Napolitano2022}, and when structures can be detected, the colours and positions of the GCs connected to them can shed further light on their origin \citep[][e.g]{Lim2017,Dabrusco2022}. Since GCs are sparse even in massive galaxies and tidal features are hard to detect, studying this connection requires deep, wide-field observations. In nearby ETGs, rich underlying substructure is related with GC systems with unusual colour and luminosity distributions \citep[e.g.][]{Sesto2016,Bassino2017}, and the presence of young GCs \citep[e.g.][]{Strader04,Woodley2010}.


The elliptical galaxy NGC\,3640, located at a distance of 26.9\,Mpc according to the results from the surface brightness fluctuation method (SBF) \citep{Tonry2001,Tully2013}, is part of a loose group conformed by approximately eight galaxies \citep{Madore2004}. This group is thought to be dynamically young since no X-ray emission was detected by ROSAT above $3\,\sigma$ of the background level \citep{Osmond2004}. NGC\,3640 has an absolute magnitude of $\textrm{M}_\textrm{B}=-20.93$ \citep{Tal2009}, and it has been classified as both E3 \citep{deVaucouleurs1991} and T3 \citep{Kormendy1982}, the latter being defined as a type for galaxies with very prominent companions, making the probability of tidal effects very high. Its close companion, NGC\,3641 is located at a projected angular distance of $2'5$ South. If we consider the distance to NGC\,3640 as the mean distance to the group, this corresponds to a physical distance of $19\,\textrm{kpc}$. It is classified as a compact elliptical (cE) in \cite{Devaucouleurs1976}, \rvw{and as a peculiar elliptical (Ep) in \cite{deVaucouleurs1991}}. \rvw{NGC\,3641 has a projected half-light effective radius of $\rm{r}_{\rm{eff}}=1.2\,kpc$ \citep{Cappellari2011} and a stellar mass of $10^{10}\,M_{\odot}$ \citep{Miller2015}, which sets it slightly offset from the relations for cEs in the literature such as \cite{Janz2016} and \cite{Ferre-Mateu2021}.} \rvw{The main properties of the galaxies according to the NASA Extragalactic Database\footnote{http://ned.ipac.caltech.edu} the Atlas3D Survey \citep{Cappellari2011} are given in Table\,\ref{tab:prop}.}

\begin{table*}[]
    \centering
    \begin{tabular}{c|c|c|c|c|c|c|}
        Galaxy & RA$^1$ & Dec$^1$ & Distance$^2$ & $M_{K}^3$ & $V/\sigma_{e}^3$ & $\rm{r}_{\rm{eff}}^3$\\
        NGC\,3640 & 11h21m06.8s & +03d14m05s & 26.9\,Mpc & -24.60 mag & 0.370 & $30.9\,\rm{arcsec}\,(1.9\,kpc)$ \\
        NGC\,3641 &  11h21m08.8s & +03d11m41s &  26.9\,Mpc & -21.85 mag & 0.331 & $9.33\,\rm{arcsec}\,(1.2\,kpc)$\\
    \end{tabular}
    \caption{Properties of both galaxies obtained from $^1$ NED, $^2$ SBF \citep{Tully2013}, and $^3$ Atlas3D Survey.}
    \label{tab:prop}
\end{table*}

\cite{Prugniel1988} (P88 for the rest of this work) present the first analysis dedicated to the morphology of this galaxy and its kinematics. Using photometry, they analyse the surface brightness profile of NGC\,3640, identifying signatures of a recent merger in the shape of the isophotes and in the residual structures. A previous work had already shown the isophotes to be boxy \citep{Bender1988}, and more recently, NGC\,3640 has even been referred to as an example of `extreme boxyness' \citep{Michard2004}. In P88, they also detect a dust lane along the minor axis, a feature that has been identified in remnants of merger processes with large quantities of gas and dust. With additional long-slit spectroscopic data, P88 characterize NGC\,3640 as a fast rotator ($V/\sigma=1.5)$, which can also be a consequence of an interaction or merging process. In comparison with other galaxies thought to be merger remnants, they argue NGC\,3640 appears to be at a more advanced stage of relaxation, though it lacks strong nuclear radio emission, which is expected to be ignited after the merging has been completed. Since NGC\,3640 shows very small amounts of gas and dust according to P88, the merger it experienced is thought to have been `dry'.

\begin{figure}[h]
    \centering
    \includegraphics[width=0.7\columnwidth]{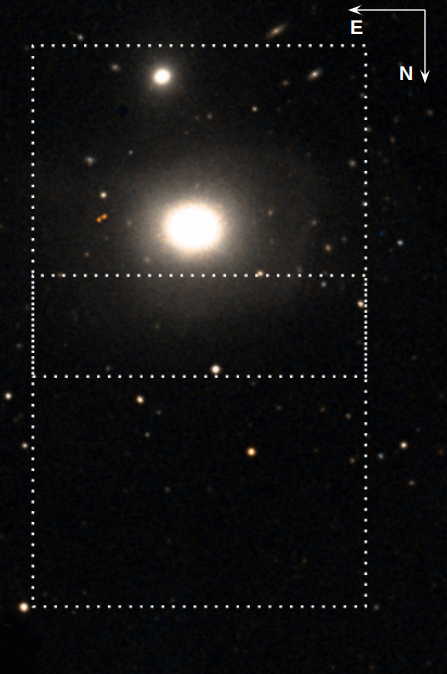}
    \caption{Representation of the observed GMOS fields plotted over a
 colored image from the Digitized Sky Survey 2, obtained from the Aladin Sky Atlas.}
    \label{fig:fields}
\end{figure}

Later on, \cite{Schweizer1992} include NGC\,3640 in a sample of galaxies they analyze using an index defined in their work to quantify features connected to mergers. Out of the 35 ellipticals in the sample, NGC\,3640 presents one of the highest values of this index, meaning it is among the most disturbed galaxies. Assuming this disturbance is owed to NGC\,3640 being the result of a disk-disk merger, they use its UBV colour to date this merger, concluding it could be a `young' elliptical, i.e. less than 7\,Gyr old. However, they point out in their work that the presence of NGC\,3641 could have had a significant influence in the disturbances detected in NGC\,3640, especially considering the high interactivity between galaxies in small groups. \cite{Michard2004} support the idea of NGC\,3640 having been formed in a major merger based on its morphology and on their analysis of its stellar population indicators, which indicate the presence of a relatively small young population, in comparison with other highly disturbed galaxies. In addition, two different works performed analysis of the stellar population of NGC\,3640 using long-slit spectra. While \cite{Denicolo2005} estimate a nuclear age of 2.5\,Gyr, the reanalysis of the same data by \cite{Brough2007} finds an older (4\,Gyr), less metal-rich center, which they estimate to be within $1.5\sigma$ of their results. Overall, their analysis points to NGC\,3640 having undergone a dissipational merger 7.5\,Gyr ago, and presenting a younger central age that would imply a more recent `wet' merger.

NGC\,3640 shows then a large number of peculiarities, including a collection of signs that point at it having undergone a relatively recent merging process. Though several of these signs have been associated with a major disk-disk merger, none of the analyses rule out interactions with the less massive galaxies in the group as the cause of some of the tidal disturbances. The compactness of NGC\,3641 could be another hint of interaction since it could be a consequence of tidal stripping. This is inconclusive since NGC\,3641 is located to the South of NGC\,3640, while the major axis of NGC\,3640 is oriented East to West, and its classification as a cE is unclear.  There are many open questions about the evolutionary history of NGC\,3640, particularly in terms of its interaction with NGC\,3641. In this work, we study the globular cluster system (GCS) of both galaxies, looking for clues that can help us untangle the interactions that have shaped NGC\,3640.

In Section 2, we describe the data and the reduction process. In Sections 3 and 4 we analyse several properties of the GCS of NGC\,3640 and NGC\,3641, respectively. In Section 5 we focus on the connection between the spatial distribution of GCs and the low surface brightness structures. In Section 6 we discuss the results. Finally, in Section 7 we summarize our results.

\section{Data acquisition and reduction}
\subsection{Observations}

\begin{table}
    \centering
    \begin{tabular}{c | c | c | c}
         Filter & $\lambda$ [nm] & Exposition time [s]  \\
         \hline
         $g'$ & 475 & 4 x 650 \\
         $r'$ & 630 & 4 x 270 \\
         $i'$ & 780 & 4 x 240 \\
    \end{tabular}
    \caption{Exposition times for each filter, identical for both fields.}
    \label{tab:exptime}
\end{table}

The observations for this work were taken utilizing the Gemini Multi-Object Spectrograph Camera (GMOS) on Gemini North (Program GN-2016A-Q-69, PI: L. Bassino) on February 10th, 2016, in imaging mode \rvw{using the g'r'i' filters with exposition times as listed in Table\,\ref{tab:exptime}.} The field of view of GMOS is of $5.5\,\rm{arcmin} \times 5.5\,\rm{arcmin}$, and we used a binning of $2\times2$, which results in a resolution of $0.146\,\rm{arcsec\, pix}^{-1}$. The seeing for these observations was $\approx 0.75\,$arcsec across all filters. We observed two fields as shown in Figure\,\ref{fig:fields}, one containing both galaxies and an adjacent one with which we aimed to cover to cover most of the extension of the GCS. \rvw{Based on the stellar mass of NGC\,3640, $1.5\times 10^{11}\,{\rm{M}_{\odot}}$ \citep{Karachentsev2015} we estimate the full extension of the GCS to be $r_{L}\approx 100\,\rm{kpc}$, i. e. $\sim12$\,arcmin according to \cite{DeBortoli2022}, although this is an upper limit, considering the environmental dependence suggested in that article for central galaxies}.

We also observed the Landolt field of standard stars SA 104 \citep{Landolt1992}, using the magnitudes obtained in the GMOS filters by \cite{Jorgensen2009}.

\begin{figure}[ht!]
    \centering
    \includegraphics[width=\columnwidth]{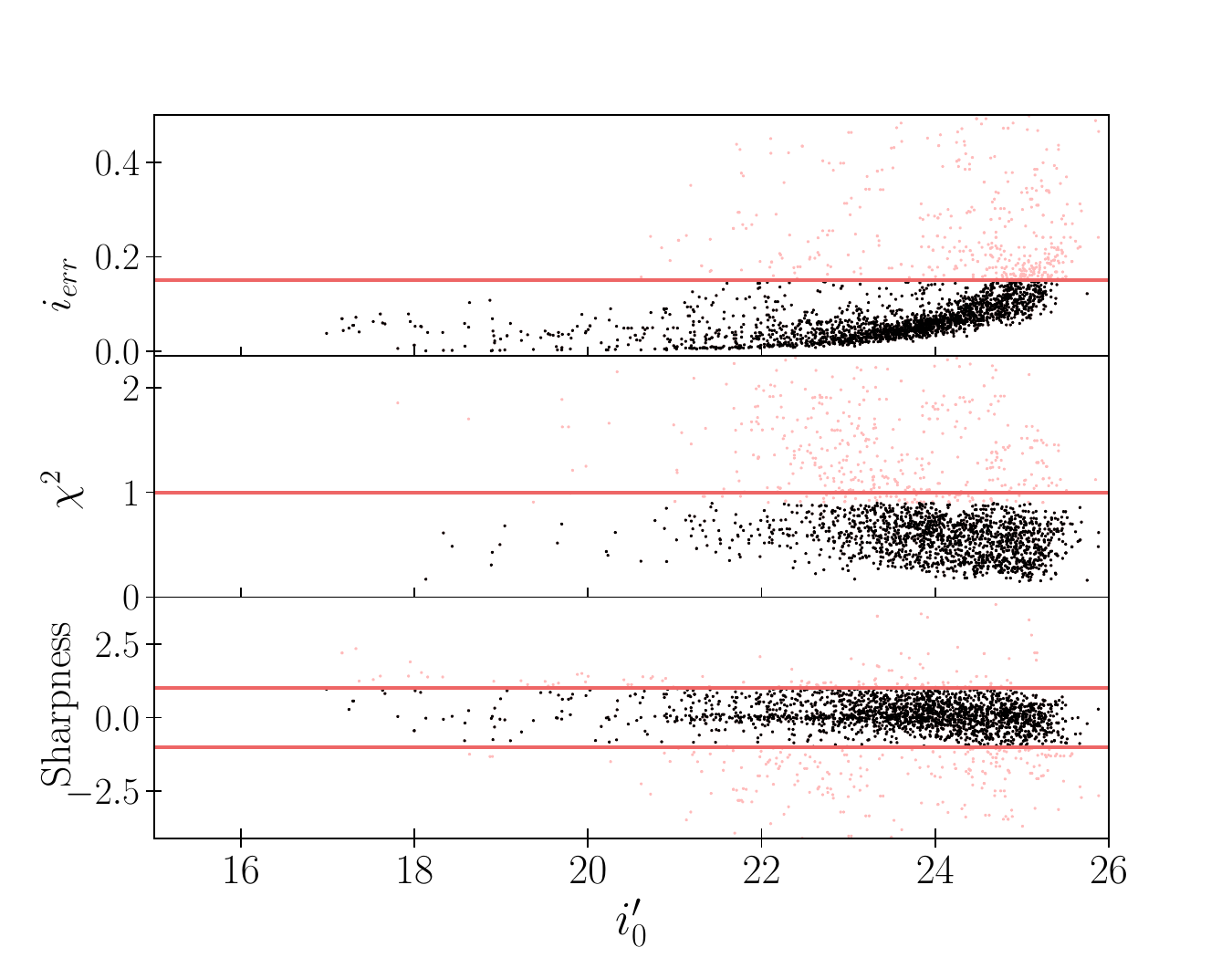}
    \caption{From top to bottom, the error in magnitude, $\chi^{2}$ and the sharpness value obtained from PSF photometry \rvw{for the $i'_{0}$ filter for reference, although the selection was carried out in all filters}. Solid lines indicate the chosen limits, with the limit in magnitude applied to the selections in the following tests.}
    \label{fig:errors}
\end{figure}

\subsection{Data reduction and point-source selection}
\label{sec:data}
We used the DRAGONS Data Reduction Software \citep{Labrie2019} to correct the observations using \textit{bias} and \textit{flat-field} images obtained from the Gemini Archive\footnote{\url{https://archive.gemini.edu/}}, and then to combine all of the images corresponding to each field in each filter. Images in the filter $i'$ presented a fringe pattern which was corrected using a blank field from the Archive. We then proceeded to subtract a model of the integrated light from both galaxies using the \textit{median filter} task in SciPy. \rvw{We ran this task two consecutive times, first with a $100\times100$ pixels window which modelled the larger structures. We subtracted this initial model from the image, and ran the task again on the residual image using a smaller window ($50\times50$\,pixels) which modeled leftover residual structures. The final image was obtained by eliminating this model from the previous residual.}


\rvw{The first catalog of objects present in the field was built using Source Extractor \citep{Bertin1996}. The run was performed twice using different filters each time, first a Gaussian filter which is considered best for detecting faint objects, and then a `mexhat' filter which is recommended for crowded fields such as the region closest to the galaxy. The resulting catalogues were combined, ensuring an efficient recovery of sources. This process was done on the three available bands. Given that the resulting catalog in both filters for the $i'$ band was the largest and contained all sources detected in $g'$ and $r'$, with a total of $\sim6000$ sources, we chose it as our base catalog at this point}. 
\rvw{Typical effective radius of a GC is $r_{\rm{eff}}=3\,$pc \citep{Jordan2007,Masters2010,Norris2014}, which at the estimated distance to NGC\,3640 corresponds to \corr{$0.03\,$arcsec. This is approximately two fifths of} the value of the seeing of these observations, resulting in GCs appearing as unresolved point-like sources. We use then the classifier $CLASS\_STAR$ within Source Extractor, which estimates the probability of the detected source being point-like. In this case, we eliminate from the sample objects with $CLASS\_STAR > 0.5$ to clear it of extended objects, mainly background galaxies and saturated stars ($CLASS\_STAR\sim0$.) }


We obtained aperture photometry for these point-like sources using the tasks in the DAOPHOT package in IRAF \citep{Stetson1987}. After selecting $\sim30$ of the brightest objects in each filter, for each field correspondingly, we built the spatially variable corresponding point-spread function (PSF) and obtained PSF photometry of our sample in the three bands with the task ALLSTAR.

\begin{figure}[h]
    \centering
    \includegraphics[width=\columnwidth]{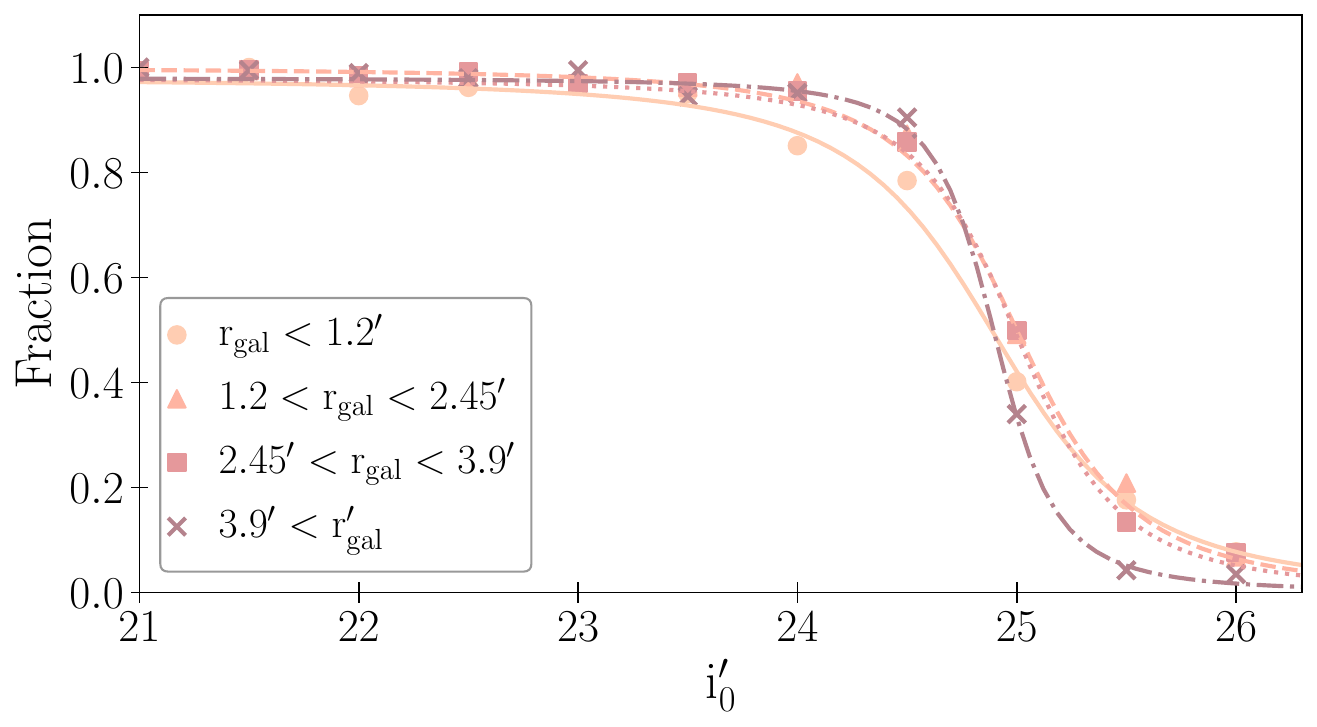}
    \caption{Completeness fraction in the $i'$ filter for \rvw{four concentric annular regions. From the innermost one outwards, they are indicated with circles and a solid line, triangles and a dashed line, squares and a dotted line, and crosses and a dot-dashed line. The magnitude is standardized as described in Section\,\ref{sec:calib}}.}
    \label{fig:comp}
\end{figure}

Besides calculating instrumental magnitudes with their errors, this task provides us with statistical parameters that characterize the goodness of the fit. In particular, it calculates the sharpness and the chi-squared for each source. In the next section, we describe the addition of artificial point-sources in the images of each band, to test the photometric completeness. Based on the parameters obtained from the PSF photometry for these artificial stars, we selected constraints for the sources from the science catalogue in each band, as shown in Figure\,\ref{fig:errors}. \corr{The constraints were analysed separately in each band, and resulted in similar enough values since the morphometric properties of the sources remained consistent across filters}. Afterwards, we match the resulting catalogues
and we obtain a total of $\sim1300$ point-like sources.

\subsection{Completeness test}
\label{sec:compl}

In the dim end of the magnitudes range, we estimate the limiting value by performing a completeness test. This process was carried out in the three bands, although we focus on the result for the $i'$ band since it was the deepest \corr{and thus the one used as a reference for the detection of sources. The detections in the $i'$ band reached $1\,\rm{mag}$ past the expected turn-over magnitude, estimated to be $i'_{0}=24.1$ as described in Section \ref{sec:lf}. In bands $g'$ and $r'$ there is a small difference in the faintest detection of approximately $0.2\,\rm{mag}$, which is expected since the integration times were chosen to achieve comparable depth.}

We build a catalog of 32000 objects, following a typical GCS radial distribution and assigning magnitudes uniformly within the expected range for colors \corr{($0.4<(g'-i')_{0}<1.5$ and $0<(r'-i')_{0}<0.5$) and magnitudes ($16<i'_{0}<26$)} of GCs in massive galaxies \citep{Harris2014}. Using the PSF obtained before, we add them as fake stars to our raw data. Then, we proceed to perform the same data reduction process as before, detecting point-like sources and performing PSF photometry in the same manner as for our \corr{sources, and applying the same selection criteria described throughout this work, including the requirement that sources be detected in all three filters}. This allows us to obtain the set of statistical parameters that return the highest fraction of sources while still providing a good fit. \rvw{We matched the detected sources to the injected ones with a maximum spatial wandering of $0.7\,$arcsec, and performed a general check to ensure the detected magnitudes did not deviate more than $0.5$\,mag from the originals.}

We divide the area into \rvw{four concentric annular regions limited by threee concentric circles centered in NGC\,3640} to take into consideration any radial variations in the completeness, and we estimate the fraction of recovered sources per magnitude bin for each of these regions. Then, we fit the function \rvw{defined in Equation 1 by \cite{Harris2009}} to the completeness fraction estimated for each region, which will be used later on to apply a more detailed completeness correction (Figure\,\ref{fig:comp}). This is particularly relevant in the inner region, in which the intrinsic brightness of the galaxy causes the completeness fraction to be lower than in other areas of the image. \corr{It is to be noted that the outermost region corresponds to the adjacent field, where completeness is lower due to slightly different observing conditions.} Finally, we determine the overall limit as the magnitude at which this fraction is of $\sim 80\%$ for the intermediate and outer regions, which in this case is $i'_{0}=24.7$. This value is thus taken as our limit in the fainter end. \corr{To check whether there is a colour dependence in the completeness levels, we split the objects at an intermediate colour, $(g'-i')_{0}=0.9$ and analysed the completeness for the two resulting samples. At the limiting magnitude, objects bluer and redder than $(g'-i')_{0}=0.9$ were $2\%$ higher and $5\%$ lower than the whole sample, respectively, showing similar differences throughout the entire magnitude range, which renders the colour dependence neglectable in terms of this analysis.}

\subsection{Photometric calibration}
\label{sec:calib}

We transform the instrumental magnitudes to the standard system by using the equations corresponding to the E2V-DD detector obtained from the Gemini Observatory website\footnote{\url{http://www.gemini.edu/}}, described in Equation\,\ref{eq:std}, and a field of standard stars, SA 104. We calculated the zero points and colour terms by fitting a curve to the relation between our photometry of the standard stars and their standard magnitudes (see Table\,\ref{tab:std}). In the case of the median atmospheric extinction coefficients, $k_{CP}$, we used the ones provided by the Gemini Observatory website, corresponding to the median value at Mauna Kea. 

\begin{equation}
\begin{split}
m_{\rm std}= & m_{\rm zero}-2.5\,\log_{10}\left(N(e-)/\textrm{exptime}\right) \\
& -k_{CP}(\textrm{airmass}-1.0)
\end{split}
\label{eq:std}
\end{equation}

\begin{table}
    \centering
    \begin{tabular}{c | c | c }
         Filter & Zero-point & Colour term  \\
         \hline
         g' & $28.08\pm0.06$ & $0.06\pm0.09*(g'-r')$ \\
         r' & $28.23\pm 0.11$ & $0.01\pm0.18*(g'-r')$ \\
         i' & $28.35\pm0.03$ & $0.13*(r'-i')$ \\
    \end{tabular}
    \caption{Zero-point and colour term values obtained for each filter. The colour term used for the $i'$ filter was obtained from the Gemini website since the quality of the observations of the standard stars field in said filter did not allow for a good fit.}
    \label{tab:std}
\end{table}

\begin{table}[]
    \centering
    \begin{tabular}{l|l}
         Statistical tests from PSF fit & $chi<0.9$  \\
        Statistical tests from PSF fit & $||sharpness||<2$ \\
         Magnitude errors & $m_{err}<0.15$ \\
         Completeness limit (faint end) & $i'_{0}=24.7$ \\
         Ultra-compact dwarf limit (bright end) & $i'_{0}=21.5$ \\ 
         Color limits & $0.4<(g'-i')_{0}<1.5$ \\
         Color limits & $0.3<(g'-r')_{0}<1.2$ \\
         Color limits & $-0.05<(r'-i')_{0}<0.5$
    \end{tabular}
    \caption{Adopted selection criteria for the GC candidates. Both the statistical tests from PSF fit and the magnitude errors were applied across all bands.}
    \label{tab:criteria}
\end{table}

\begin{figure}[ht!]
    \centering
    \includegraphics[width=\columnwidth]{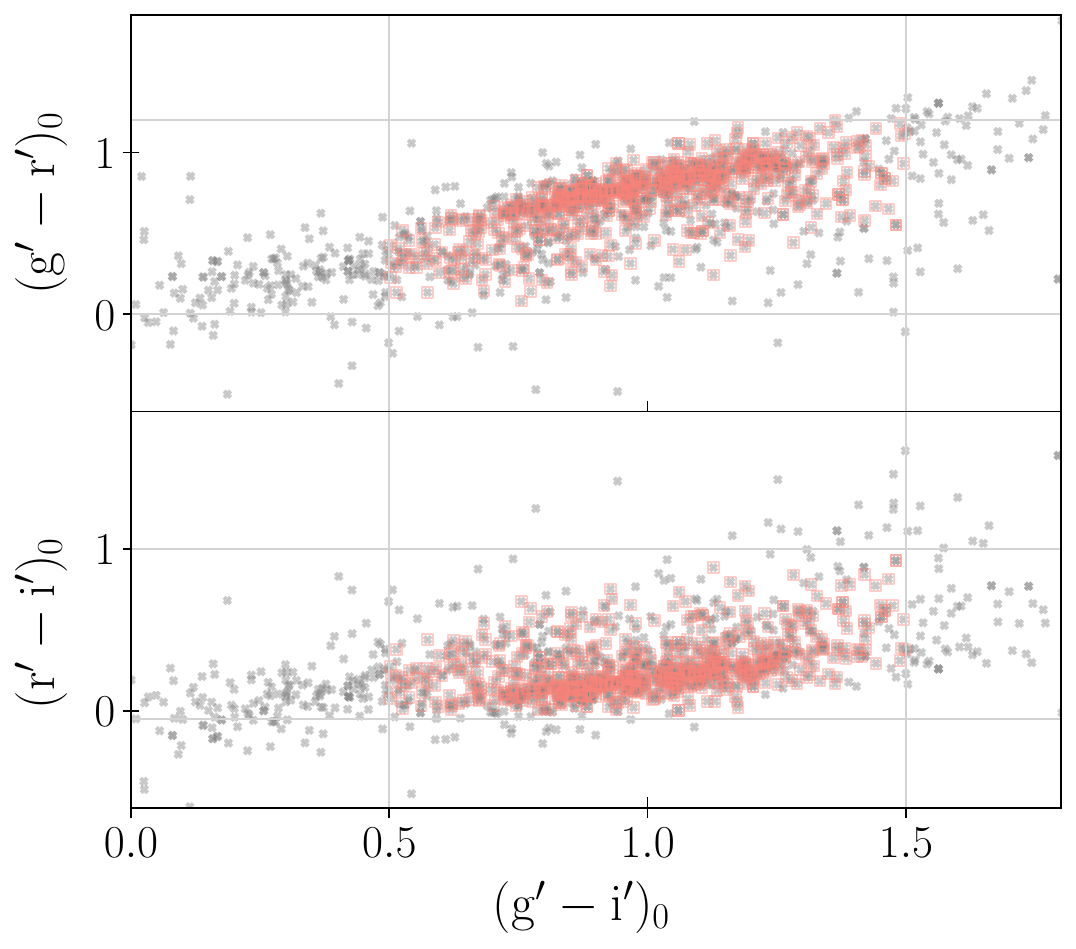}
    \caption{Colour-colour diagrams for two different combinations of filters, showing all
    sources present in the field as grey crosses, and those that satisfy all the criteria to be considered GC candidates are highlighted with pink squares. The limits shown in solid lines are 
    estimated based on values from the literature.}
    \label{fig:dcc}
\end{figure}

\subsection{GC candidate selection}

\begin{figure*}[ht!]
    \centering
 \includegraphics[width=\columnwidth]{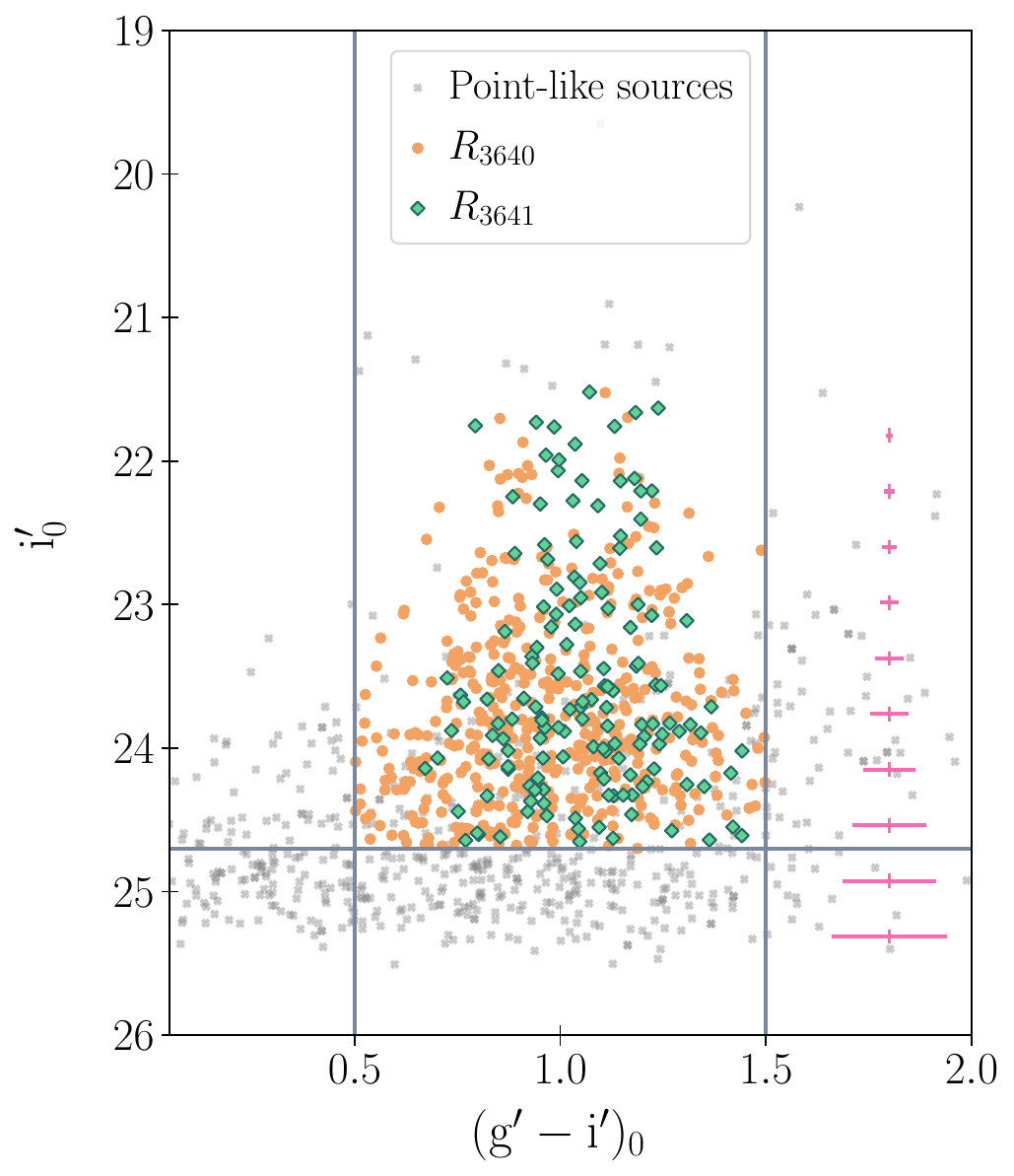}
 \includegraphics[width=\columnwidth]{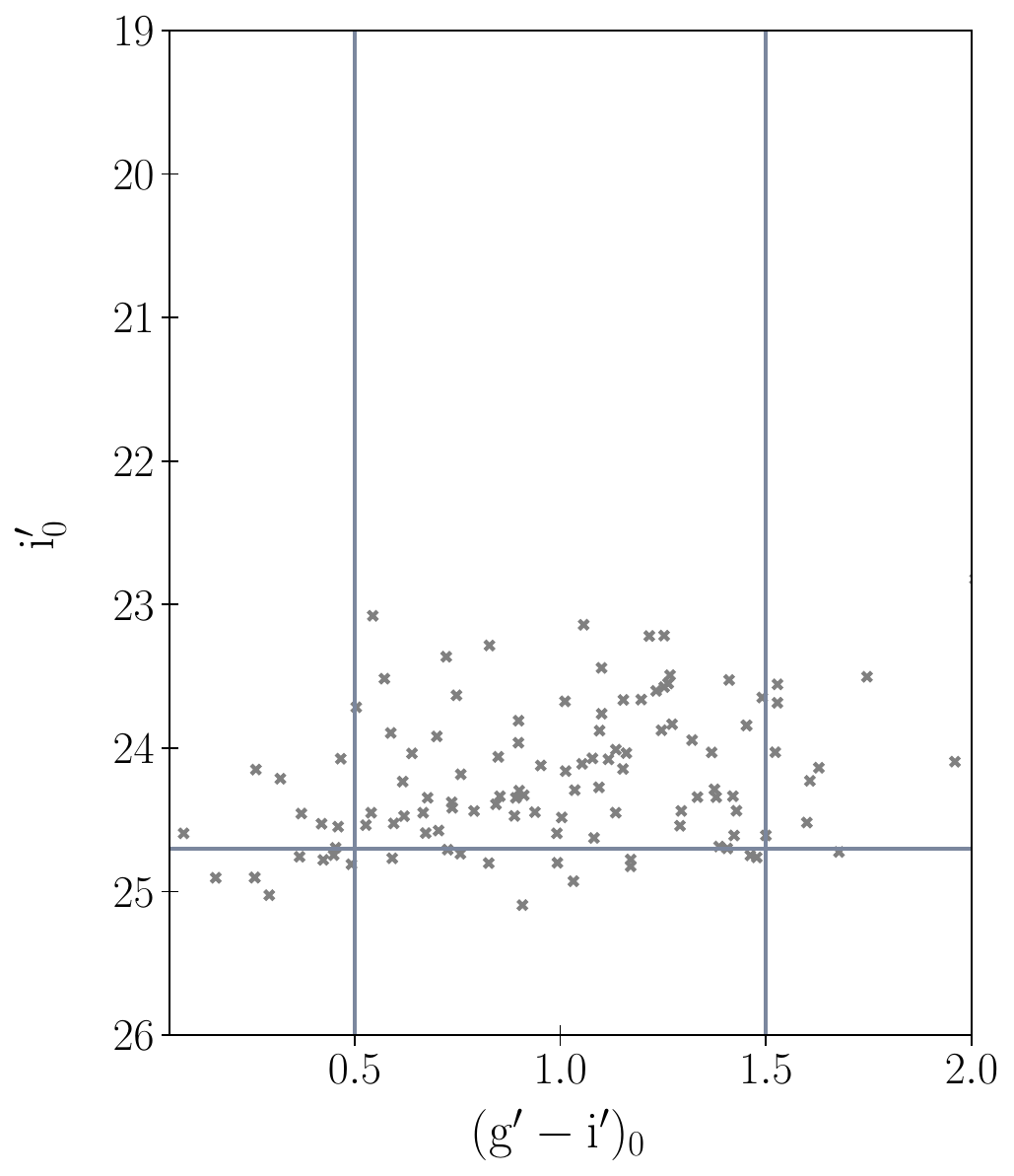}
    \caption{\textbf{Left}: Colour-magnitude diagram for all sources, shown as grey crosses. The GC candidates attributed by galactocentric distance to NGC\,3641 are shown as green diamonds, while those attributed to NGC\,3640 are represented by orange dots. In pink we show the error bars for the mean error in magnitude bins ($0.4$\,mag). \textbf{Right}: Colour-magnitude diagram for background sources.}
    \label{fig:dcm}
\end{figure*}

In Figure\,\ref{fig:dcc}, we show the colour-colour diagrams of the entire sample of point-like sources detected in the field. \corr{We use colour limits to identify GC candidates  in the three classic colour combinations, applying the same cuts as in \cite{Bassino2017}}. 
The limits in $(g'-i')$ are shown in the colour-magnitude diagram (Figure\,\ref{fig:dcm}), where we add the limit obtained from the completeness test, corresponding to $80\%$, in the dim end of the magnitudes axis. Across all bands the limits correspond to $0.4<(g'-i')_{0}<1.5$, $0.3<(g'-r')_{0}<1.2$ and $-0.05<(r'-i')_{0}<0.5$\,mag. In addition, we limit the error in magnitude to $<0.15$\,mag in all filters. As shown in \cite{Harris2014} objects brighter than $M_{V}=-11$ found around massive ETGs are more likely to be ultra-compact dwarfs, and the probability of finding GCs becomes very low. As mentioned in the Introduction, the SBF distance to NGC\,3640 is $26.9$\,Mpc \corr{(i.e. (m-M)=32.15)}, resulting in a cut-off magnitude on the brighter end of ${i'_{0}}=21.5$. \corr{After applying the criteria based on the statistical tests from the PSF fit to the catalog that resulted from matching the sources detected in all three bands, we performed a new selection using these cuts in colour and magnitudes}. All selection criteria are summarized in Table\,\ref{tab:criteria}. This is the final catalog used for the analysis of the GCS, \corr{consisting of 725 objects}.


\begin{figure}[ht!]
    \centering
    \includegraphics[width=\columnwidth]{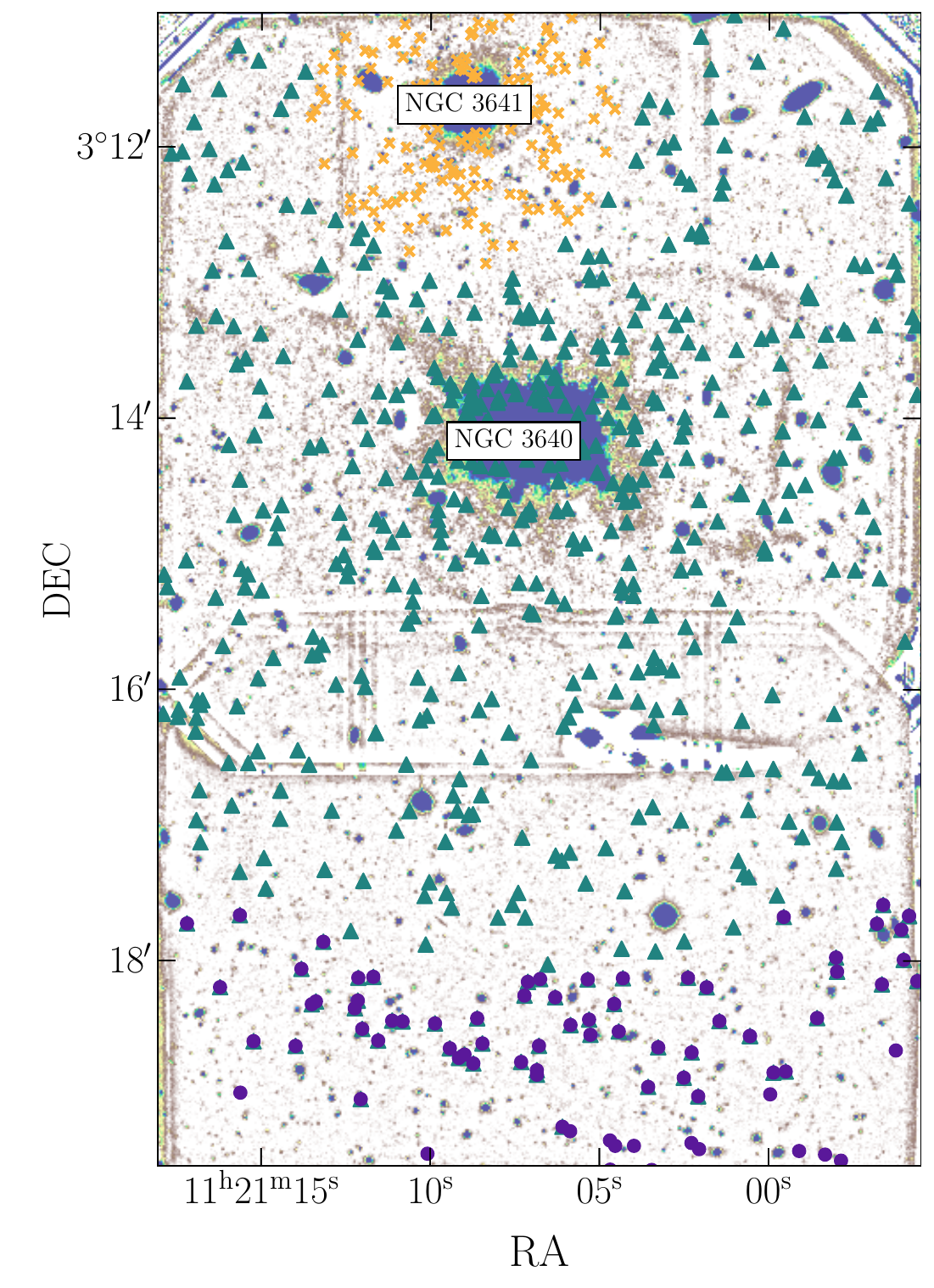}
    \caption{Spatial distribution of all GC candidates across both fields. Orange dots and green diamonds indicate sources attributed to NGC\,3640 and NGC\,3641 respectively. \rvw{Purple circles indicate objects within the colour ranges used for GCs considered background sources.} The centre of each galaxy is shown in large symbols with thick edges.}
    \label{fig:espa_dots}
\end{figure} 

Since we cannot estimate radial velocity measurements from photometry alone, initially we separate GC candidates according to their spatial distribution, as shown in Figure\,\ref{fig:espa_dots}. The relation between the stellar mass of the galaxy and the extension of the GCS shown in Equation\,9 in \cite{Caso2019} results in a value of $\sim21\,$kpc (i.e. $\sim2.6$\,arcmin at the assumed distance) for the extension of the GCS of NGC\,3641, \rvw{considering a stellar mass of $10^{10}\,{\rm M_{\odot}}$ \citep{Miller2015}. The extension according to the effective radius of the galaxy, however, is $\sim6\,$kpc (i.e. $\sim0.8\,$arcmin) following the relation by \cite{Kartha2014}. Considering this and the distance to NGC\,3640 which creates a limitation to how far we can extend the analysis, } we initially attribute all GC candidates within a radius of $1.25$\,arcmin of the center of NGC\,3641 to its GCS, and the rest to NGC\,3640, though this separation is by no means conclusive.

\section{The GCS of NGC\,3640}

\subsection{Colour-magnitude diagram}

In Figure\,\ref{fig:dcm} we show the colour-magnitude diagram of the point-like sources in the field and the GCs associated with each galaxy. NGC\,3641 presents intermediate colours in comparison with NGC\,3640, covering a smaller range in general. In particular, NGC\,3641 almost lacks blue GCs. \rvw{If the galaxies have interacted, it is reasonable to expect blue GCs initially from NGC\,3641 to have been accreted first by NGC\,3640, considering that blue GCs have more extended distributions than red ones \citep[e.g.][]{Strader2011,Kluge2023}. These accreted blue GCs may now be mixed with those of NGC\,3640 or they could present a much more extended spatial distribution which is limited in this analysis by the hard boundary set to separate both systems.}

NGC\,3640 covers a colour range typical of GCS in massive ETGs, and towards the brighter end, two subpopulations are evident. 

The right panel shows the background sources, located in the region selected as background based on the radial distribution described in the following section. The contamination level is even across the entire colour range.

\subsection{Radial distribution}

\begin{figure}[ht!]
    \centering
    \includegraphics[width=\columnwidth]{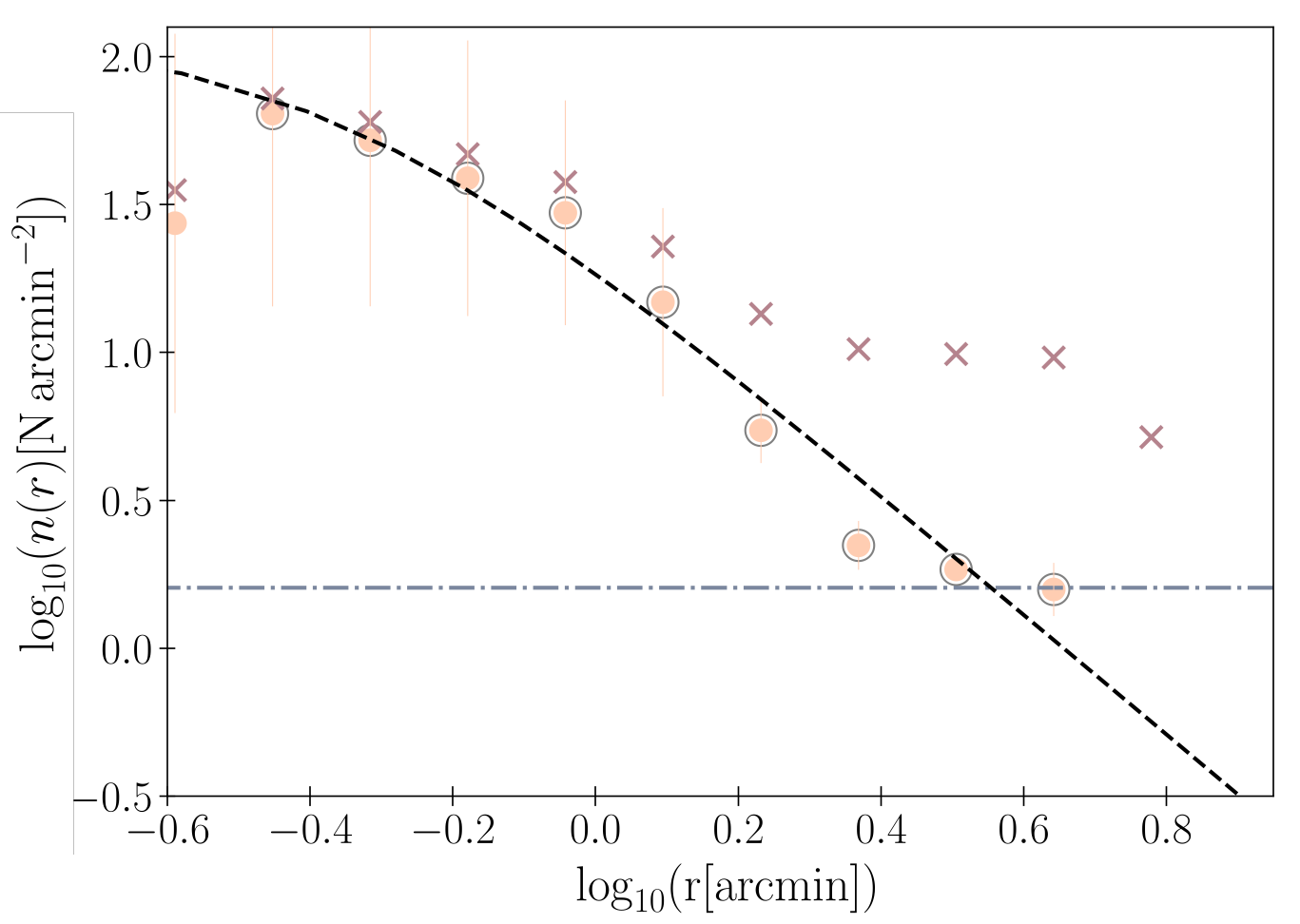}
    \caption{Radial distribution for GC candidates attributed to NGC\,3640. Crosses describe the raw sample, while orange circles show the background corrected distribution, with a grey, larger circle indicating those considered for the fit. The dashed line shows the level at which $20\%$ of the background density is reached, and the dash-dotted curve, the Hubble-Reynolds law fit. }
    \label{fig:drad}
\end{figure}

In Figure\,\ref{fig:drad} we show the radial distribution, corrected by completeness, for the GC candidates across both fields attributed to NGC\,3640. We estimate the background level considering the flattening of the distribution in the outer regions, at $r_{\rm{gal}}=3.7\,$arcmin, obtaining a contamination correction of $8\pm3\,\textrm{GC}\,\textrm{arcmin}^{-2}$ which was applied throughout the analysis.

We fit a Hubble-Reynolds law to the background corrected profile, resulting in the following profile:

\begin{equation}
\label{eq:hubble}
n(r) [N \rm{arcmin}^{-2}]= 1.88 \left(1+\left(\frac{r [\rm{arcmin}]}{1.16}\right)^2\right)^{-2.06}
\end{equation}

\corr{Previous studies of GCS on wide fields of view have shown that adopting a projected density of 0.2-0.3 times the background density results in consistent measurements of the extension of the GCS \citep[e.g.][]{Bassino2006,Caso2017}}. We extrapolate then this function to the value at which it reaches $20\%$ of the background level, and calculate a total extension of the system of $31\,{\rm kpc}$ ($\approx 4\,$arcmin). 

Integrating the profile up to this point, we obtain the total population up to our completeness level, resulting in $268\pm9\,$GCs.

\begin{figure}[ht!]
    \centering
    \includegraphics[width=\columnwidth]{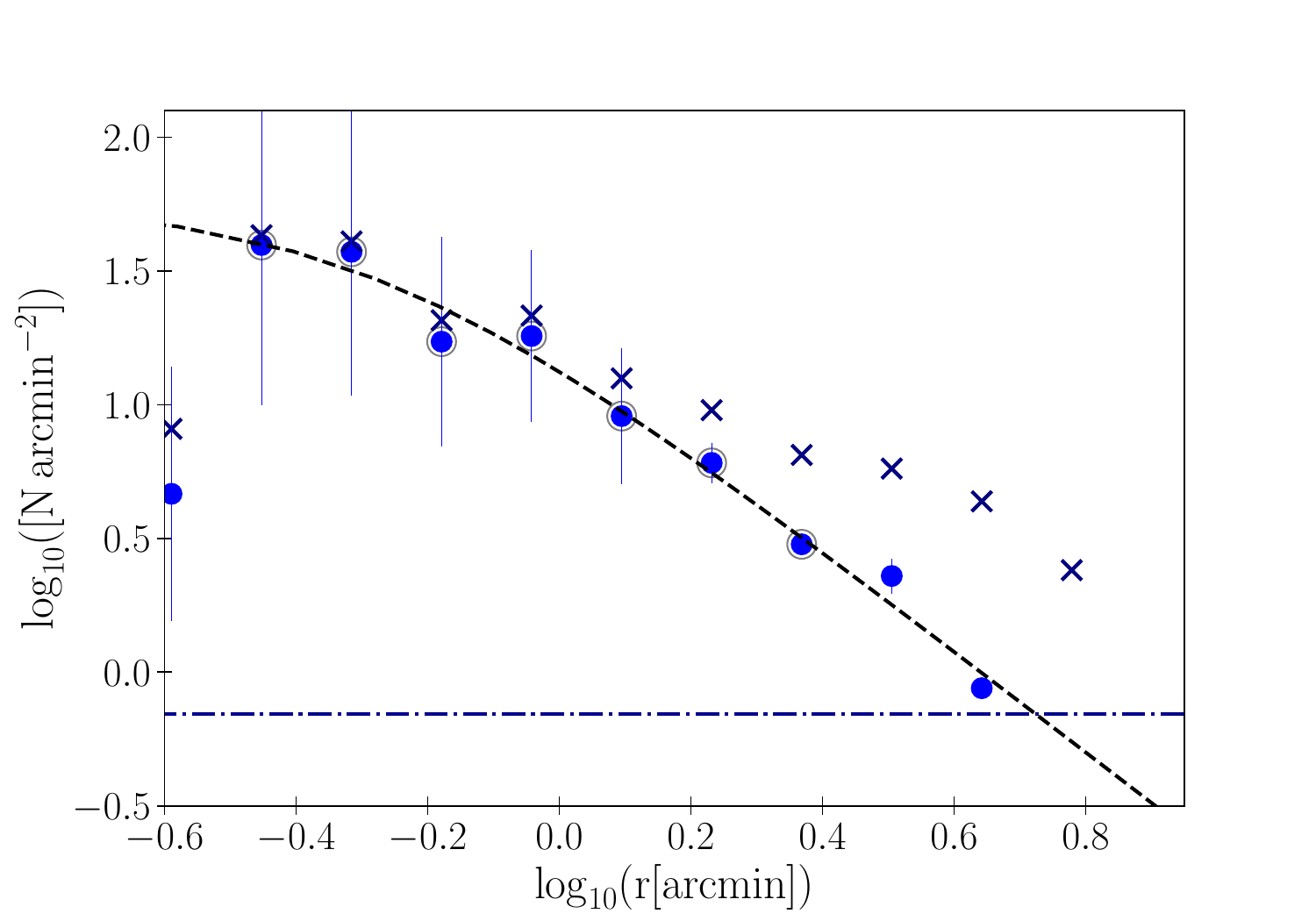}
    \includegraphics[width=\columnwidth]{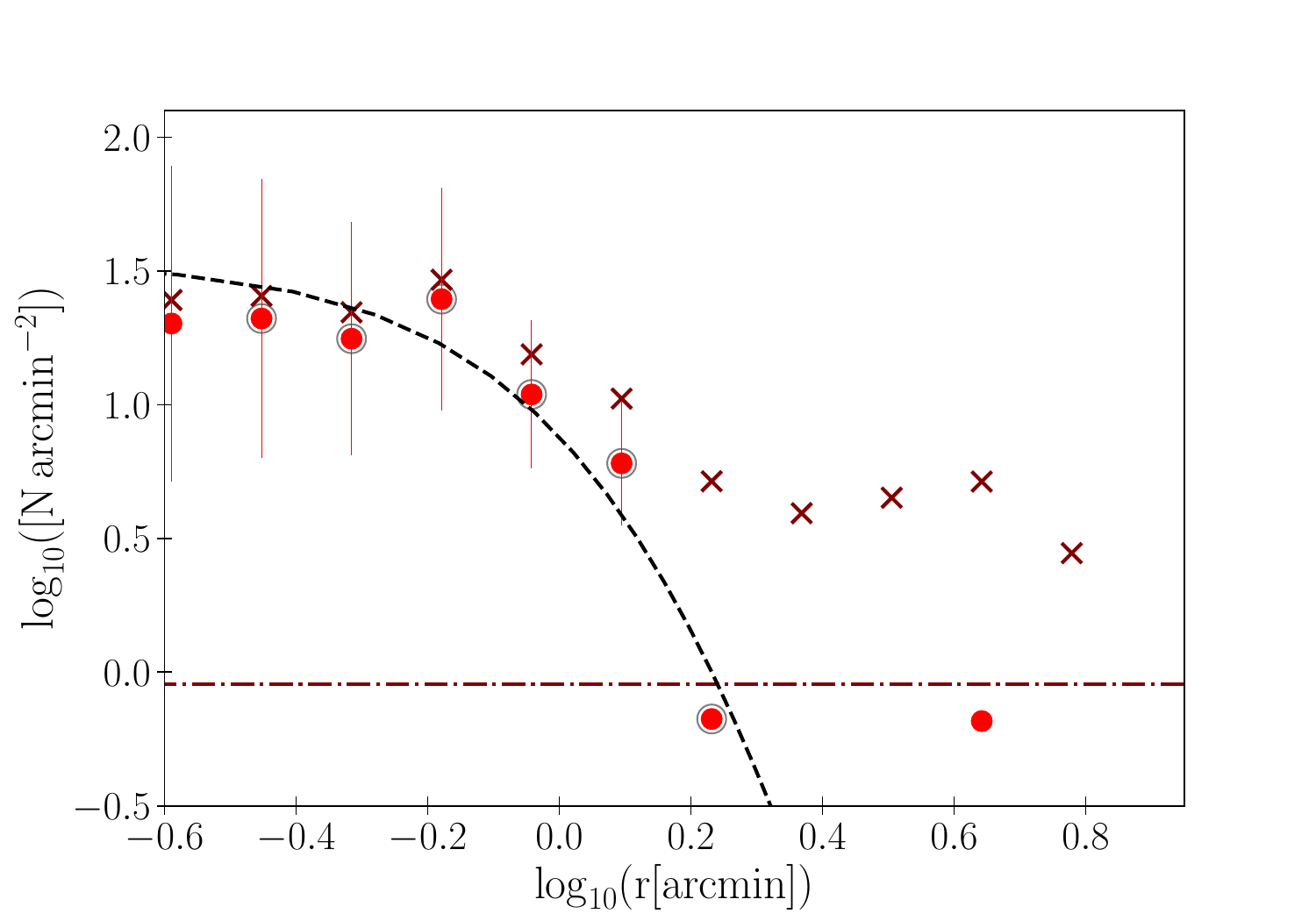}
    \caption{Radial distribution for GC candidates attributed to NGC\,3640, for the blue (top) and red (bottom) subpopulations. Crosses describe the raw sample, while circles show the background corrected distribution. The dash-dotted horizontal line indicates $20\%$ of the background level, and the dashed curve, the Hubble Reynolds law.}
    \label{fig:dradcol}
\end{figure}

In addition, in Figure \ref{fig:dradcol}, we show the radial distribution for GC candidates split into red and blue subpopulations, following the separation described in the following Section. We fit Hubble-Reynolds laws to both profiles, obtaining core radii of $0.72$\,arcmin for red GCs and $1.8$\,arcmin for blue GCs. The red subpopulation appears heavily concentrated towards the centre of the galaxy, whereas the blue one extends to the outer region in a steeper manner. These behaviours are consistent with what is expected if we consider red GCs to be mostly formed in-situ, and blue GCs to come from accretion processes that increase their density in the outer halo \citep[e.g.][]{Reina-Campos2019,El-Badry2019,Harris2023}.

\subsection{Colour distribution}

\begin{figure}[ht!]
    \centering
    \includegraphics[width=\linewidth]{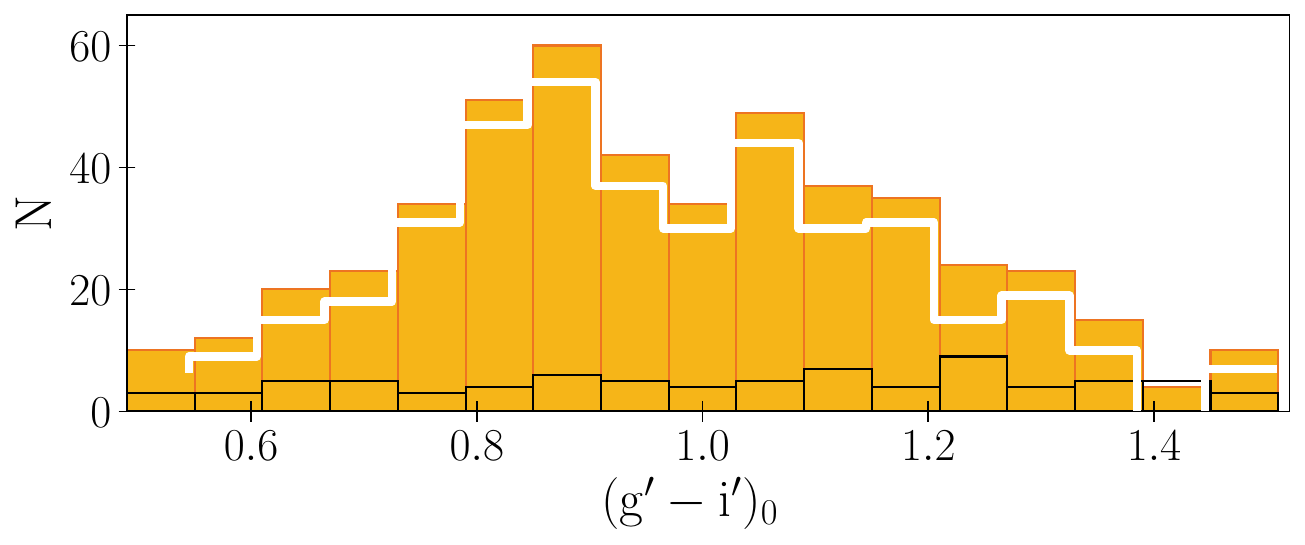}
    \caption{Colour distribution for the full sample of GC candidates attributed to NGC\,3640 shown in orange. Black lines represent the background distribution, and white lines the corrected distribution. }
    \label{fig:dcol}
\end{figure}

In Figure\,\ref{fig:dcol} we show the histogram of the colour distribution for the GCs attributed to NGC\,3640, as well as the background correction. \rvw{This correction is estimated by scaling the amount of sources in the background region to the area considered for the GCS.} Since the background sources show a homogeneous distribution across the colour range, we consider the background correction negligible in terms of fitting the subpopulations. 

\begin{figure}[ht!]
    \centering
    \includegraphics[width=\linewidth]{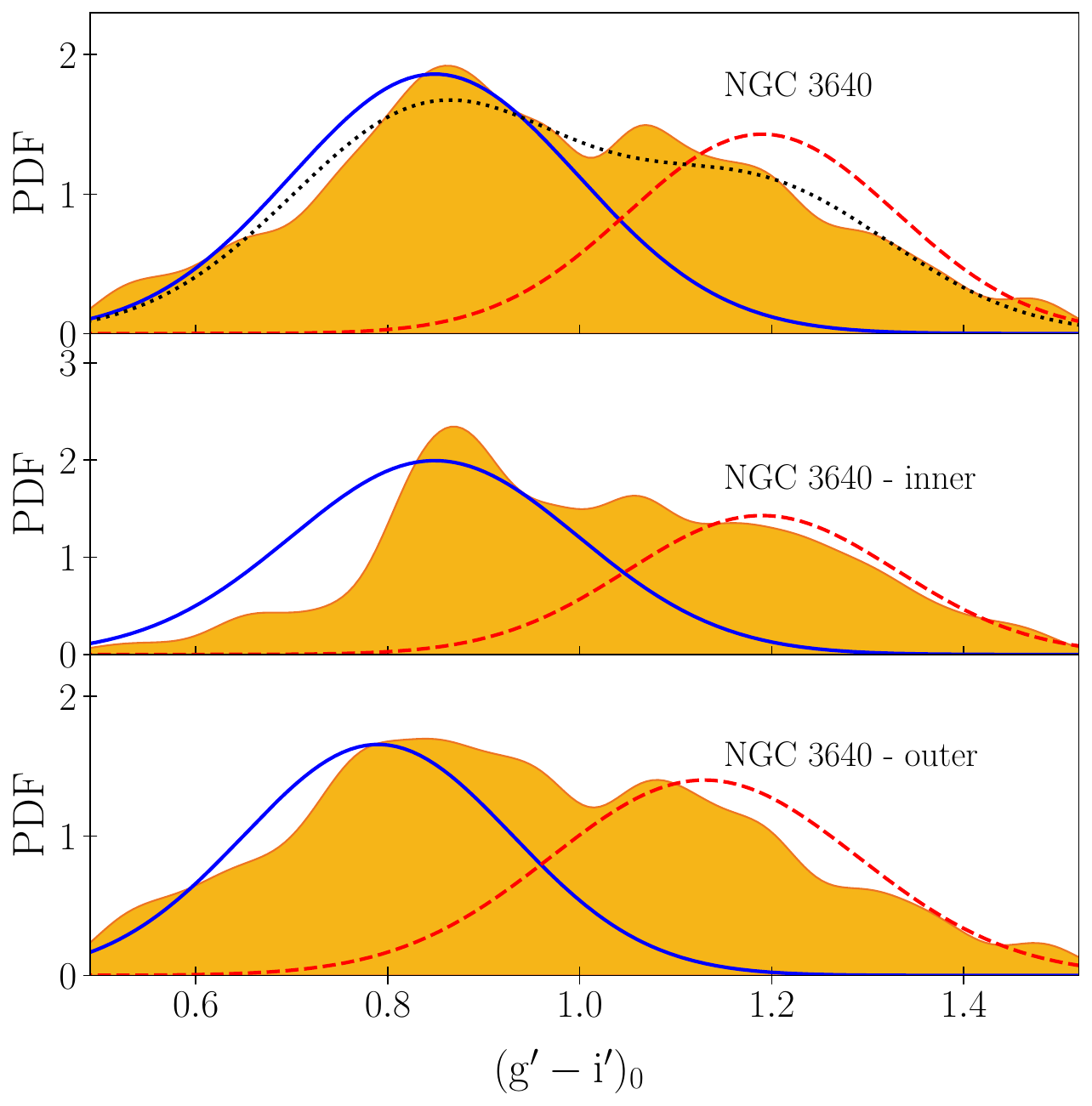}
    \caption{In orange curves, smoothed colour distributions for the full sample of GC candidates attributed to NGC\,3640 (top). In the middle panel, the inner region (left), and in the bottom, the outer region of NGC\,3640. Blue solid curves and red dashed curves represent the best fit from GMM for bimodal Gaussian distributions for each sample.}
    \label{fig:dcol2}
\end{figure}

Figure\,\ref{fig:dcol2} shows the density probability functions of the colour distribution for the GCs attributed to NGC\,3640, as well as the inner ($r_{gal}<1.25\,$arcmin) and outer ($r_{gal}>1.25\,$arcmin) regions. \corr{The distance to separate the regions was estimated by identifying an abrupt drop in the distribution of red GCs at around this value, which could be an indication of a change in the dominating physical process. Slight variations were considered and found to have no significant impact.}. These smoothed distributions use a Gaussian kernel, with a bandwidth of $0.5$.

We use the Gaussian Mixture Modeling code (GMM, \cite{Muratov2010}) to estimate the probability of the distribution being bimodal in all samples. The output provided by GMM includes both the parameters for the best fit of Gaussian functions ($\mu$ and $\sigma$, as well as the fraction of sources attributed to each mode) and statistical tests that estimate the confidence level of the multi-modal fit. These statistical tests are the kurtosis of the distribution, which quantifies its peakedness, and D, which is a parameter that is calculated by assuming bimodality and measuring the distance between the estimates means, weighed by the broadness of each mode. D is expected to be greater than 2 in bimodal distributions. A unimodal distribution would result in a stronger peak, which would in turn be reflected in positive values for the kurtosis.

In Table\,\ref{tab:gmm}, we show the results for each sample. When considering the entire population of GCs attributed to NGC\,3640, the result is a negative kurtosis and a $D>2$, meaning a bimodal distribution, with modal values similar to those in other massive early-type galaxies \citep{Harris2009,Forbes2011}. 

Both regions are statistically bimodal according to this fit. \corr{Both subpopulations present redder means in the inner region in comparison to their means across the whole field, with the effect being more noticeable for red GCs.} The outer region presents bluer means for both subpopulations than the entire sample, and is more likely to be contaminated by GCs connected to NGC\,3641.

\begin{table*}     
    \centering
    \begin{tabular}{|c|c|c|c|c|c|c|c|c|}
        \hline
        \textbf{Sample} & \multicolumn{2}{c|}{\textbf{$\mu$}} & \multicolumn{2}{c|}{\textbf{$\sigma$}} & \textbf{Red fraction} & \textbf{Kurtosis} & \textbf{D}\\
        \hline
        & \textbf{Blue} & \textbf{Red} & \textbf{Blue} & \textbf{Red} & & & \\
        \hline
        NGC $3640$ & $0.85\pm0.04$ & $1.19\pm0.05$ & $0.15\pm0.02$ & $0.14\pm0.02$ & $0.35$ & $-0.56$ & $2.27\pm0.24$ \\
        NGC 3640 (inner) & $0.92\pm0.05$ & $1.24\pm0.05$ & $0.14\pm0.03$ & $0.12\pm0.02$ & $0.30$ & $-0.56$ & $2.54\pm0.39$ \\
        NGC 3640 (outer) & $0.79\pm0.05$ & $1.13\pm0.06$ & $0.14\pm0.02$ & $0.16\pm0.02$ & $0.48$ & $-0.60$ & $2.22\pm0.29$ \\
        \hline
    \end{tabular}
    \caption{Parameters for the Gaussian functions resulting from the bimodal fit performed using GMM for the different samples.}
    \label{tab:gmm}

\end{table*}

\subsection{Luminosity function}

\label{sec:lf}

The luminosity function of globular cluster systems is widely used as a distance estimator due to the universality of the absolute magnitude of its turn-over \citep[e.g.][]{Villegas2010,Rejkuba2012}, \rvw{with a value of $M_V=-7.4$.} The SBF distance to NGC\,3640 is $26.9$\,Mpc ($m-M=32.15$), which results in a turn-over magnitude of $V_{0}=24.75$. Using the transformations from \cite{Faifer2011} shown in Equation\,\ref{eq:faifer}, and a mean $(g-i)$ value of 1, we can transform to approximately $i_{0}=24.1$\,mag, which is well within our completeness range, allowing us to estimate our own measurement of the distance to compare with the literature value. 

\begin{equation}
    V_{0} = g_{0} - 0.34(\pm 0.02)(g-i)_{0}-0.03(\pm0.02)
\label{eq:faifer}
\end{equation}

\begin{figure}[ht!]
    \centering
    \includegraphics[width=\columnwidth]{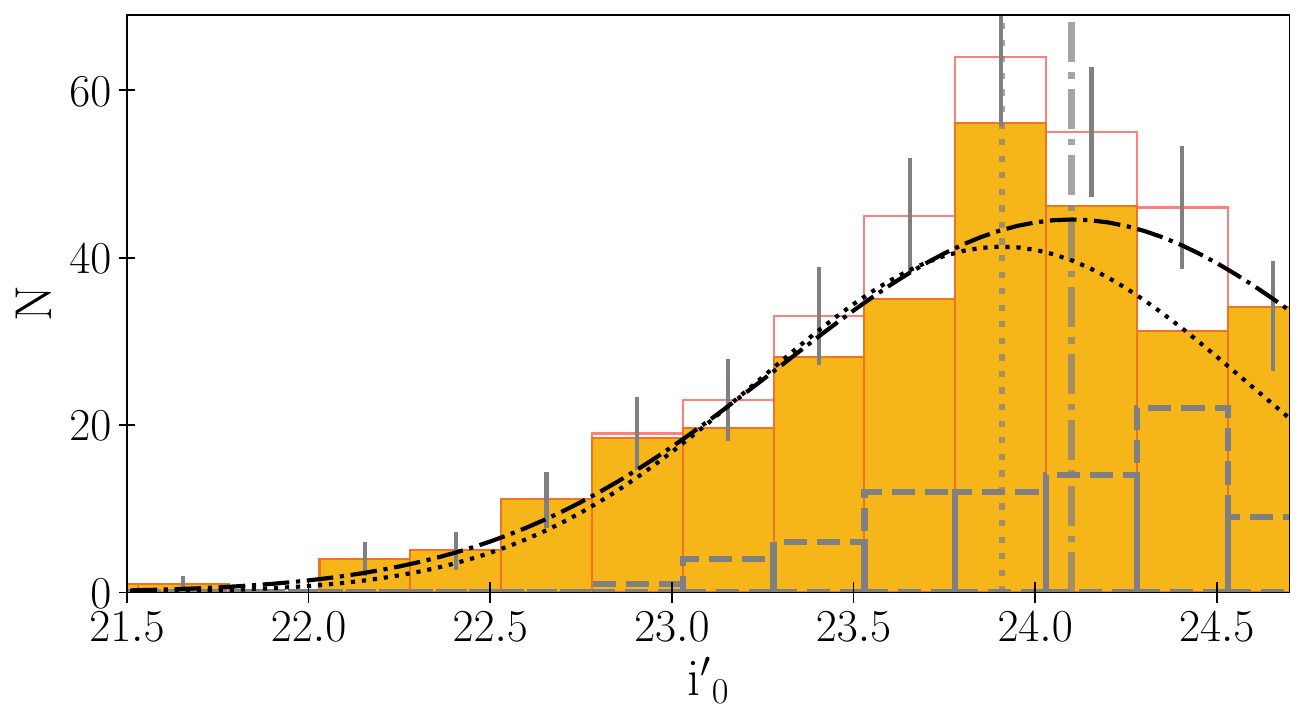}
    \caption{Luminosity function for the GC candidates associated with NGC\,3640. Orange lines show the original distribution, with orange solid blocks showing the result after correcting for completeness and subtracting the background, indicated in dashed gray lines. The dashed curve and vertical line correspond to the fit performed in this work and its corresponding turn-over magnitude, while the dash-dotted is the fit obtained with the distance from the literature, with its fixed turn-over magnitude shown as a dash-dotted vetical line. All fits were done using only the shown magnitude range according to the completeness test, with the corrected distribution.}
    \label{fig:lf}
\end{figure}

In Figure\,\ref{fig:lf}, we show the luminosity function for GC candidates associated with NGC\,3640. \corr{It is derived in the $i'$ band since the photometric uncertainties were slightly lower than in the other two.}
This distribution was corrected using the completeness functions shown in Section\,\ref{sec:compl}, accounting for radial variations. A background correction was also applied, considering a scaling to the area analysed.. Fitting a Gaussian function to our distribution results in a turn-over value of $i_{0}=23.9\pm 0.2$\,mag, where the error was obtained through bootstrapping. \rvw{The distance modulus for this galaxy according to the SBF distance quoted before is $32.1\pm0.1$\,mag, which results in a turn-over magnitude of $i_{0}=24.1\pm0.1$\,mag}. The dispersion of the Gaussian is $\sigma_{i}=0.67\pm0.2$\,mag. In \cite{Harris2014}, the GCLF is proposed to have an intrinsic width of $\sigma_{L}=1-1.25\,\textrm{mag}$ for a galaxy of this stellar mass. The lack of complete angular coverage in the GMOS field and the contamination from NGC\,3641 might be causing this difference.

We use the turn-over value from SBF and the dispersion of the Gaussian function fit \corr{to our distribution} using this fixed value, $\sigma_{L}=0.81$\,mag (slightly larger than the one obtained in our fit) to estimate the percentage of GCs within our completeness range in relation to the total population, which in this case is $74\%$. Combining this result with the number obtained from the integration of the radial distribution, we can estimate the total number of GCs for NGC\,3640, $N_{GC}=362\pm9$. \corr{This is made possible by the three passbands having the same depth at the turn-over magnitude.}

\rvw{Considering the stellar mass of NGC\,3640, we can calculate the parameter T as defined by \cite{Zepf1993}, which is the number of GCs per stellar galaxy mass normalized to $10^{9}M_{\odot}$. We obtain $T=2.3\pm0.1$. Comparing to Figure\,5 in \cite{DeBortoli2022}, which compares $T$ values to the stellar mass of the host galaxy, this is within the expected values for a galaxy of this mass.}


\section{The GCS of NGC\,3641}

\subsection{Radial distribution}

\begin{figure}[ht!]
    \centering
    \includegraphics[width=\columnwidth]{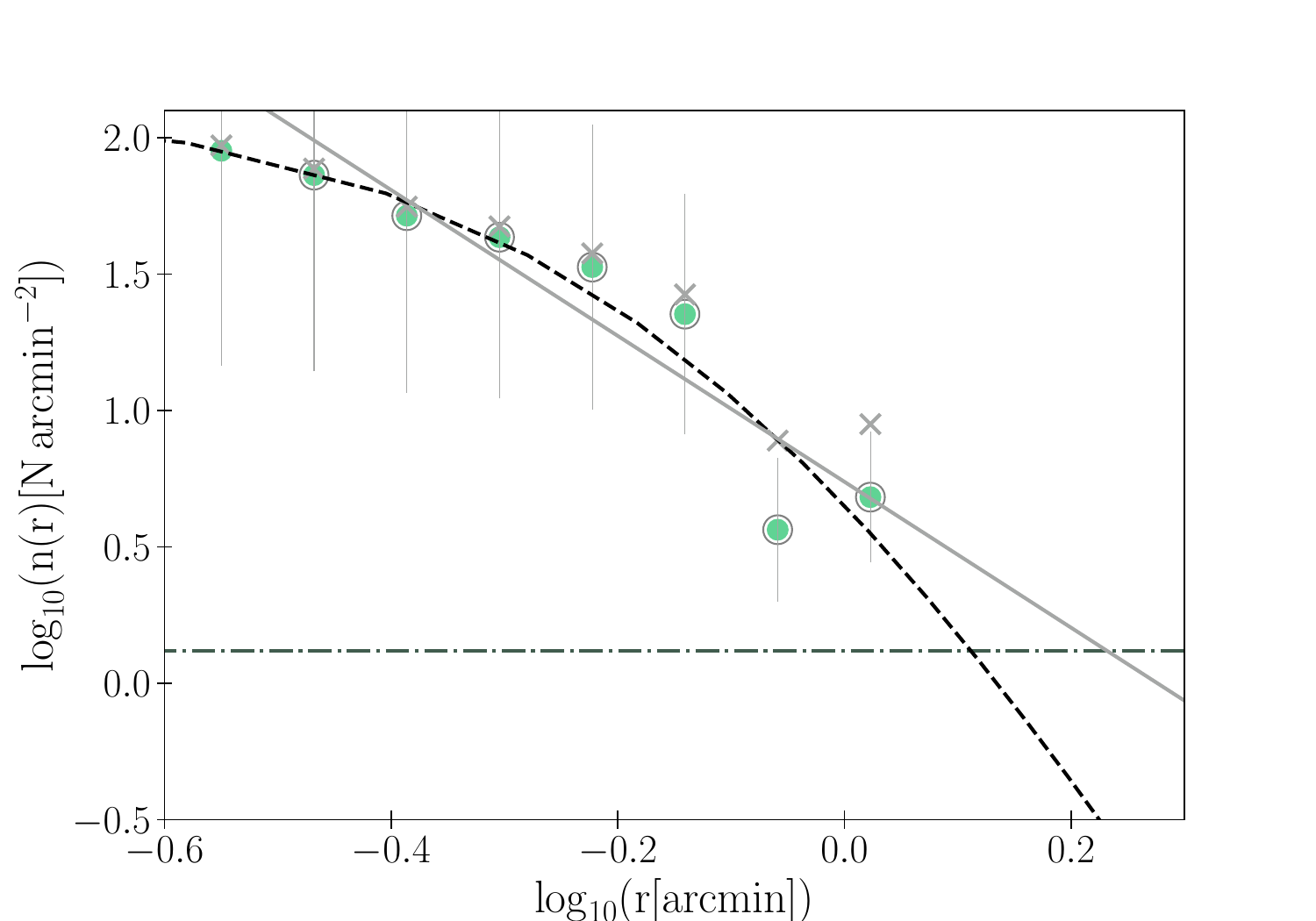}
    \caption{Radial distribution for GC candidates attributed to NGC\,3641. Grey crosses describe the raw sample, while green dots show the background corrected distribution. A dashed curve shows the Hubble profile. A horizontal line indicates the same background level as in previous figures.}
    \label{fig:dradc}
\end{figure}

In Figure\,\ref{fig:dradc} we show the radial distribution for the GCs within $1.25\,$arcmin of the centre of NGC\,3641 corrected by completeness and background, with a Hubble-Reynolds law fit up to the imposed limit, \rvw{resulting in:}

\begin{equation}
\label{eq:hubble2}
n(r) [N \rm{arcmin}^{-2}]= 2.11 \left(1+\left(\frac{r [\rm{arcmin}]}{0.69}\right)^2\right)^{-2.99}
\end{equation}

If we extrapolate this profile to the same level we used for NGC\,3640, we obtain an estimate for the population of NGC\,3641 up to the completeness level of $101\pm25$\,GCs.

\subsection{Colour distribution}

\begin{figure}[ht!]
    \centering
    \includegraphics[width=\columnwidth]{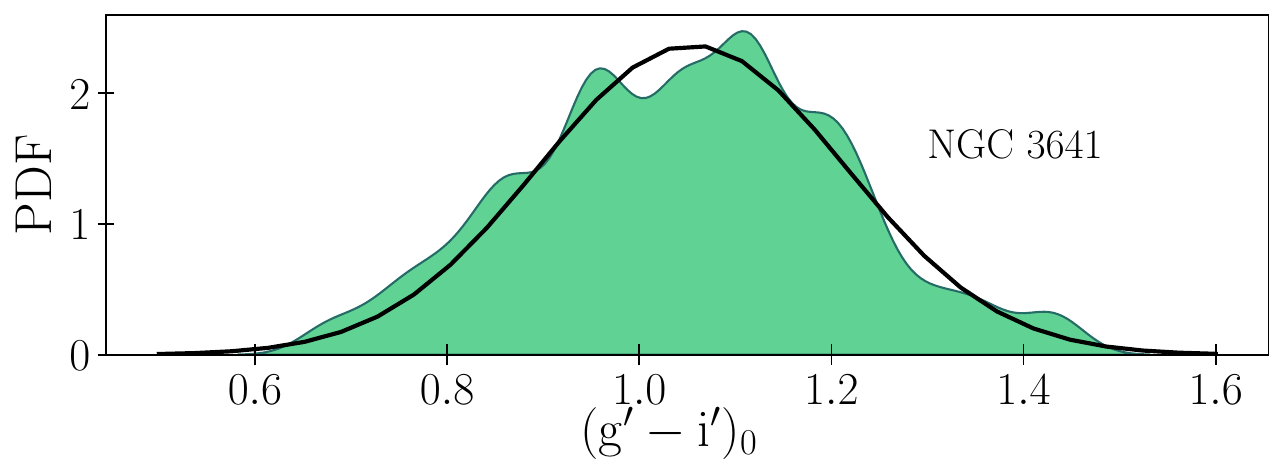}
    \caption{Smoothed colour distribution for GC candidates associated to NGC\,3641. A black curve indicates the unimodal distribution resulting from the GMM fit.}
    \label{fig:coldist41}
\end{figure}

In Figure\,\ref{fig:coldist41} we show the colour distribution for GCs associated with NGC\,3641. Analogously to the previous section, we use GMM and in this case obtain a negative kurtosis ($-0.384$) but the value of D is significantly smaller than 2 ($1.74$), indicating an unimodal distribution is a better fit. The mean for the distribution is $(g'-i')_{0}=1.05$\,mag, with $\sigma=0.16$\,mag. In the bimodal fit, the fraction of red GCs is $0.84$, with a mean of $1.09$, showing most of the GCs surrounding NGC\,3641 can be considered red. This is reasonable considering we are only analysing objects really close to the galaxy, where red GCs are expected to be dominant, and also because if NGC\,3640 is actually stripping its GCs, it would have stripped mainly blue ones which are more external.

\subsection{Luminosity function}

\begin{figure}[h]
    \centering
    \includegraphics[width=\columnwidth]{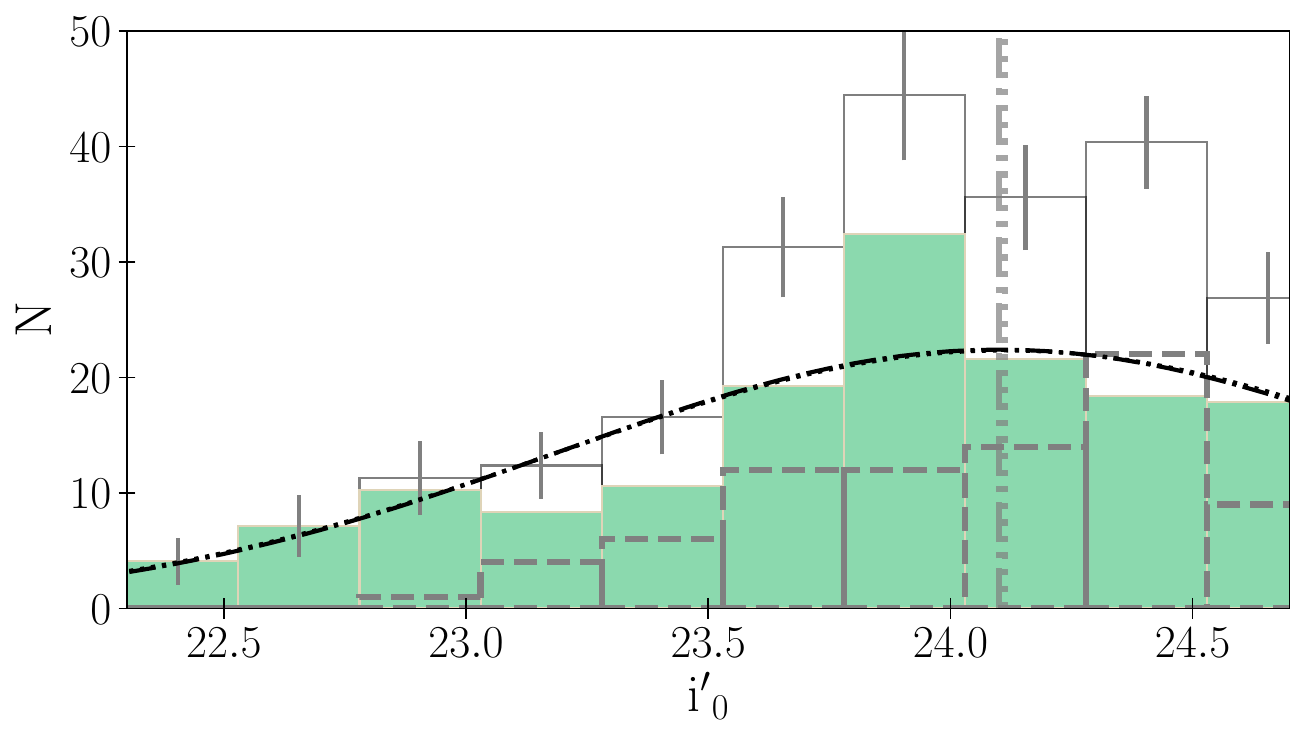}
    \caption{Luminosity function for the GC candidates associated to NGC\,3641. Dashed grey lines show the background distribution, green solid bars indicate the result from subtracting it from the raw distribution (white bars with grey solid lines). The dot-dashed curve shows the Gaussian fit.}
    \label{fig:lf1}
\end{figure}

In Figure\,\ref{fig:lf1} we show the luminosity function of GC candidates associated to NGC\,3641, corrected by completeness and background estimation. We perform a similar analysis as with NGC\,3640, performing the fit of a Gaussian function both using the turn-over from the literature (in this case, the same one as for NGC\,3640), and then fitting our own value. We find an almost exact match, $i'_{0}=24.11\,\pm0.3$, although our luminosity function is noisy, with a dispersion \corr{slightly smaller} than usual ($0.9$\,mag) compared to the intrinsic width mentioned above from \cite{Harris2014}, and showing bumps towards the brighter end which may be due to contamination from NGC\,3640.

\rvw{In this case, the GCLF indicates that our coverage up to our completeness limit is of $82\%$. Considering the number of GCs estimated from the radial distribution, the mean value corresponding to the SBF distance and the dispersion estimated from our distribution, we calculate a total population of $120\pm25$ GC.}

\section{Spatial distribution and tidal features}
\begin{figure*}[ht!]
    \centering
    \includegraphics[width=0.8\linewidth]{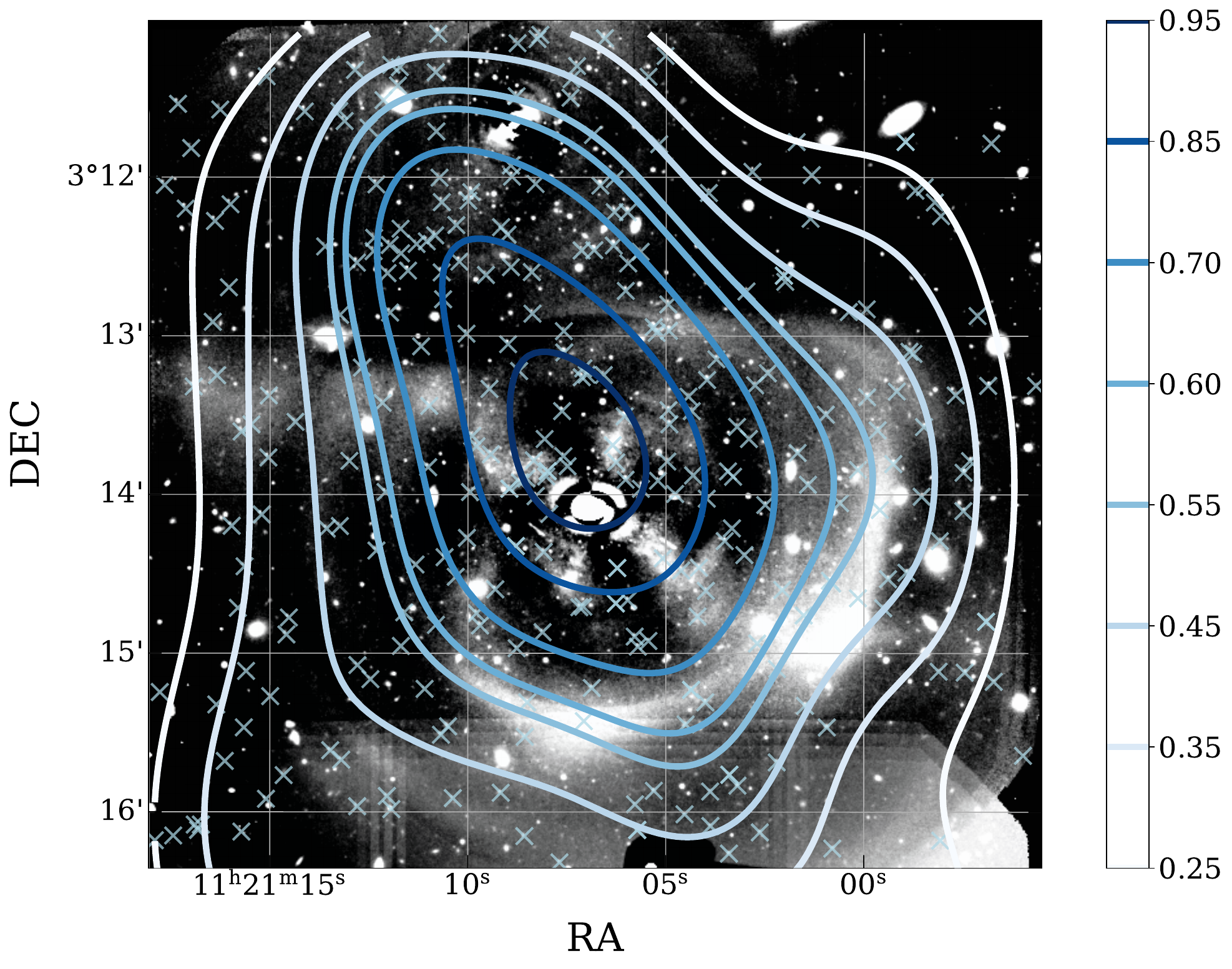}
    \includegraphics[width=0.8\linewidth]{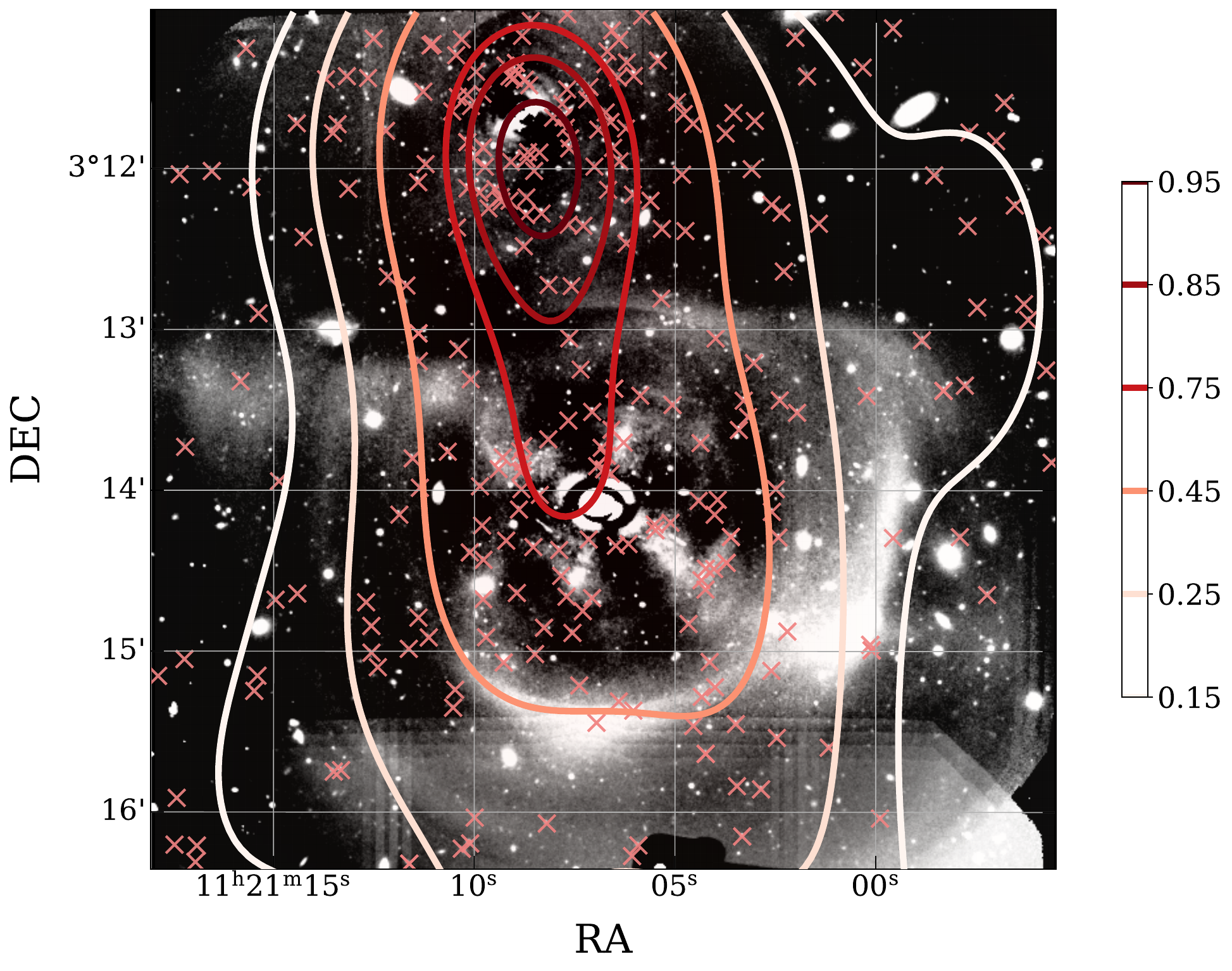}
    \caption{Smoothed distribution of GC candidates across the entire field separated between blue (top) and red (bottom), with density contours plotted in solid lines. Crosses in both panels show the individual positions, and the background is the residuals image of the field obtained in the $i'$ filter after modelling both galaxies with the task ELLIPSE and subtracting them from the raw image.}
    \label{fig:spa}
\end{figure*}

The panels of Figure\,\ref{fig:spa} show the contours of the spatial distribution of the GC candidates located across the field containing both galaxies, separated by colour, placing the limit in $(g'-i')=1$\,mag, drawn on top of residual images. These were obtained modelling both galaxies using the ELLIPSE task in iraf, and subtracting it from the image. \rvw{For the contours we used a 2D kernel density estimate using a Gaussian kernel, using the sklearn package \citep{scikit-learn}.}

In the top panel, we can see that the blue GCs are extended across the entire field. The contours follow one of the strongest substructures, identified by P88 as `A'. This behaviour may indicate that the interaction that caused the shell-like structure in `A' also pushed the blue GCs in this direction.

\corr{Whereas blue GCs present overdensities surrounding the galaxies, and seem to be missing from the region closest to NGC\,3641, the red GCs around NGC\,3641 show an asymmetric distribution, with an overdensity towards NGC\,3640 that can be interpreted as a bridge that connects them. This seems to indicate that NGC\,3640 is accreting GCs from its neighbour, and this process is ongoing.}

\begin{figure*}[ht!]
    \centering
    \includegraphics[width=0.8\linewidth]{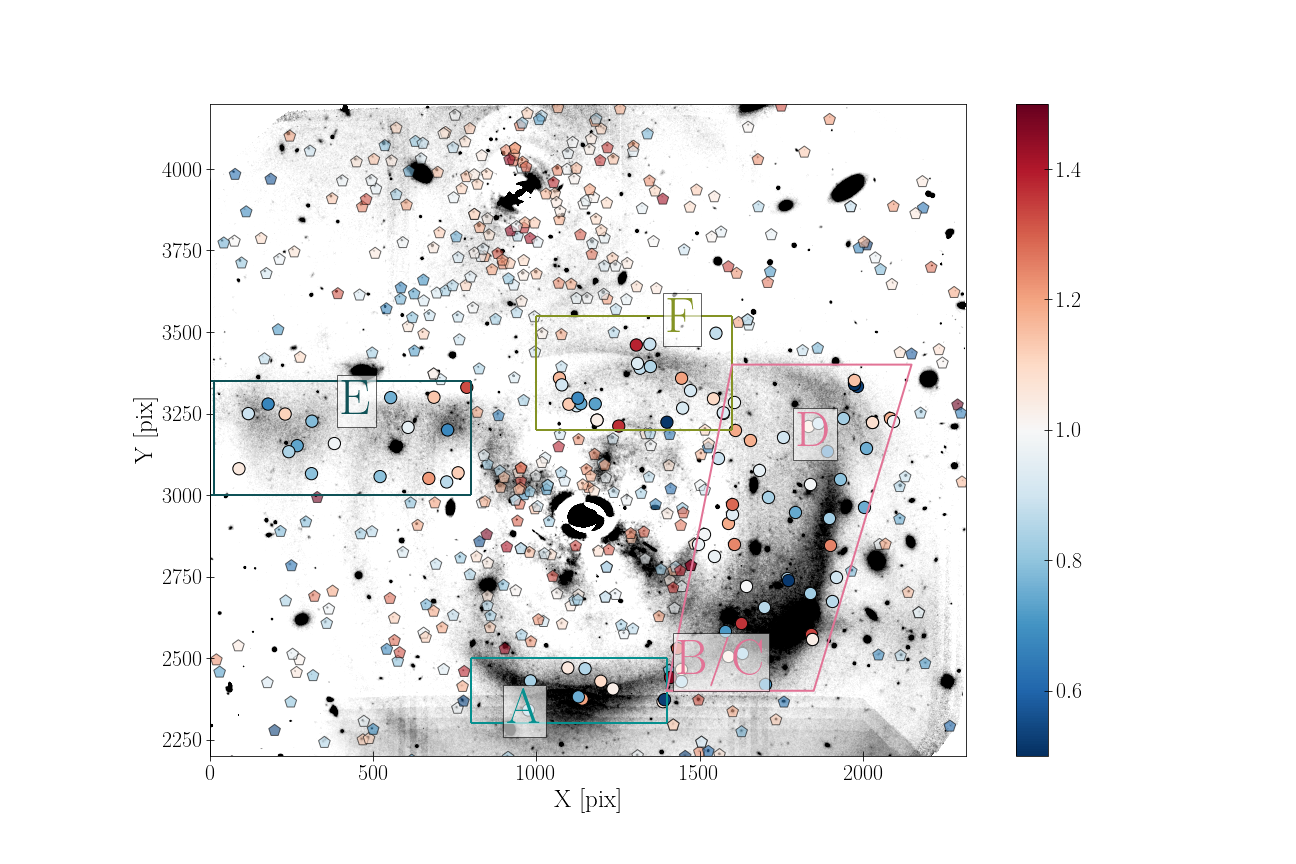}
    \caption{Image in the $i'$ filter resulting from subtracting the model of the surface brightness profile of both galaxies. Circles indicate GC candidates across the entire field with their colours shown in a gradient, and polygons indicate the location of the shells following the nomenclature given in P88.}
    \label{fig:shells}
\end{figure*}

\begin{figure}[ht!]
    \centering

    \includegraphics[width=\columnwidth]{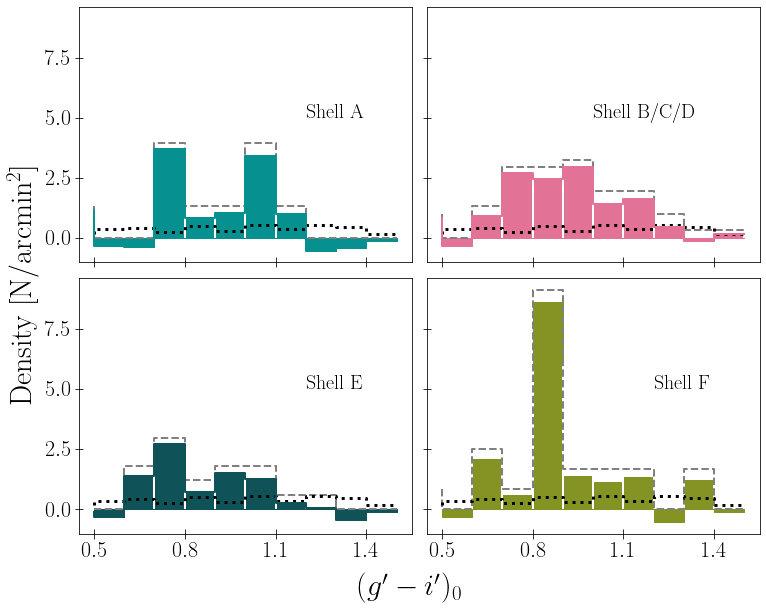}
    \includegraphics[width=\columnwidth]{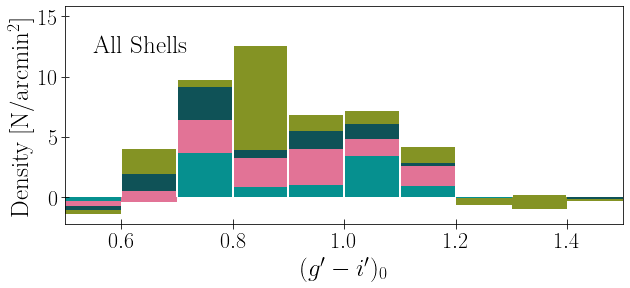}
    \caption{Histogram of GCs located in the regions defined as shells. Top image shows each shell separately, bottom one shows the combination of all the histograms, following the same color scheme as in the polygons in the previous Figure.}
    \label{fig:shells_abcd}
\end{figure}

\section{Discussion}

\subsection{GCs as tracers of substructures}

In the current paradigm of ETG evolution, both major and minor mergers play important roles at different stages of galaxy evolution. Gas-rich major mergers dominate the early stages, while minor mergers become more relevant during the second phase of evolution, where ETGs grow their mass by accretion with very low to null rates of star formation \citep{Oser2010, Naab2007,Hilz2012}. These encounters leave traces behind as different types of substructures, e.g. streams, shells, tails, and plumes \citep{Mancillas2019}[and references within]. We identify the substructures in NGC\,3640 as shells since they are approximately semi-circular and concentric, mostly located on the northwest corner of the galaxy. \cite{Mancillas2019} attributes the fact that shells tend to accumulate on one side of the galaxy to the consequences of a satellite accretion event, where the arc is formed after the satellite falls into the central potential. In general, shells are expected to form after mergers with a 4:1 mass ratio at most, i.e. intermediate- to major-mass mergers \citep{Pop2018,Karademir2019}, and they survive around 4\,Gyr. This upper limit for the timing of the accretion event is consistent with previous analyses of NGC\,3640 which date the structures and parts of the stellar population (see Section \ref{sec:intro}).

The mergers the host galaxy undergoes shape the GCS. They may increase the population by either forming new GCs if there is enough gas involved and if the conditions are violent enough, or accreding old GCs from satellite galaxies. They may also alter their spatial distribution, redistributing them to the halo \citep{Choksi2019,Dornan2023}. 

\cite{Dabrusco2022} study statistically significant structures in the spatial distributions of the GCS of a variety of Fornax galaxies, concluding they are relics of accreted GC systems that are overlaid on the smooth distribution of the GCs belonging to the host galaxy. Following their analysis, we show in Figure\,\ref{fig:shells_abcd} the colour distribution for the GCs that correspond spatially to the shells detected in the galaxy, identified in Figure\,\ref{fig:shells}. In shell F, located in the direction of NGC\,3641, we find most GCs have a similar color, around $(g'-i')_{0}=0.9$\,mag, bluer than those associated to NGC\,3641, \rvw{where the mean colour was found to be $(g'-i')_{0}=1.05$\,mag}. In substructures A and E, blue GCs dominate, while the region covered by substructure B/C/D has a wider color distribution, although it is also dominated by blue objects, including GCs with colors bluer than $(g'-i')_{0}=0.7$\,mag. 
 
This analysis is based on the distribution observed in 2D being representative of the physical position of the objects in the real, 3D Universe. \cite{Dabrusco2022} argue that the risk of the 2D structures being merely projection effects is small. This is due to previous studies like \cite{Woodley2011} showing through simulations that the probability of misidentifying a 2D subgroup is $1\%$, and to the fact that the identified structures present complex morphologies that a smooth 3D spatial distribution cannot explain.

\subsection{Paucity of GCs in the inner region}

\begin{figure}
    \centering
    \includegraphics[width=0.5\linewidth]{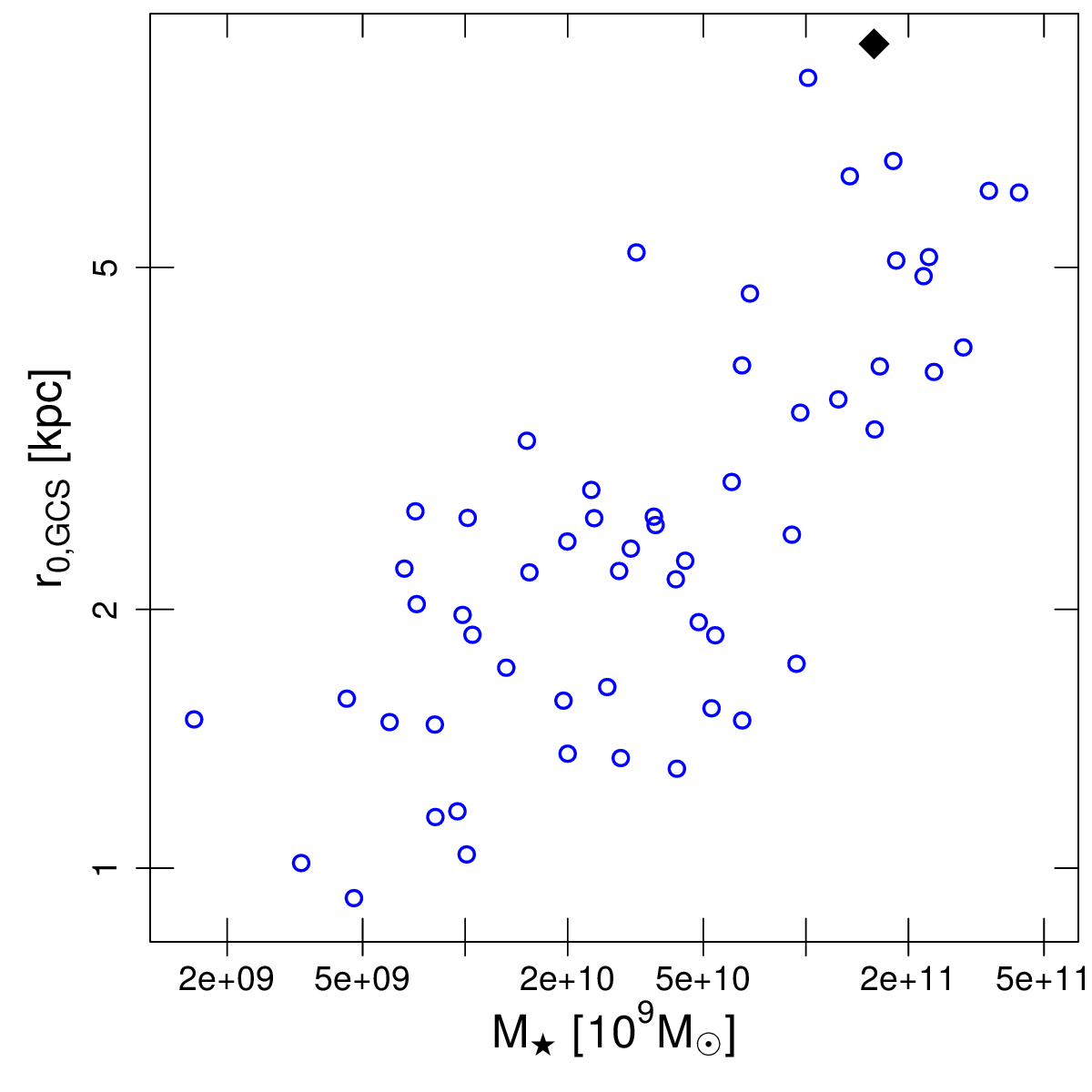}
    \caption{Core radius of the GCS vs. stellar mass of the galaxy for the sample in \cite{Caso2024} in blue circles. A black diamond indicates the position of NGC\,3640.}
    \label{fig:caso}
\end{figure}

The Hubble-Reynolds profile fit to the projected radial distribution of the GCS of NGC\,3640 has a core radius of $r_{0}=1.16\,\textrm{arcmin}=9.07\,kpc$. Considering a stellar mass of $log{M_{\star}}=11.2$ we can place it in the context of the $r_{0}$ vs. $M_{\star}$ relation studied in \cite{Caso2024} for a sample of early-type galaxies in HST/ACS archival date (see Figure\,\ref{fig:caso}). Given that Gemini data is not comparable in terms of depth with HST/ACS, even after a thorough completeness analysis, this number might be slightly enlarged in comparison to those in the sample in \cite{Caso2024}, \rvw{and GC detection is typically less efficient in the inner regions of galaxies}. However, it is clear NGC\,3640 has a larger core radius than most galaxies in the sample, especially for those with similar masses. This implies the GCS of NGC\,3640 is flatter up to a larger galactocentric radius, which can also be seen by simple visual inspection of the projected spatial distribution in Figure\,\ref{fig:shells}, where the paucity of GCs near the centre of the galaxy is apparent. 

The highest core radius value in the HST/ACS sample \citep{Caso2024} corresponds to NGC\,4406 (M86), a galaxy that presents an offset peak in its GCS connected to a potential disturbance caused by a satellite dwarf galaxy, and a shell that is both seen in the low-surface brightness analysis \citep{Mihos2017} and the GC spatial structure \citep{Lambert2020}. The two following galaxies are NGC\,1316 and NGC\,4472 (M49), both having undergone late, major mergers. Numerical simulations of hierarchical galaxy formation including GCs have shown that mergers or large accretion events such as being tidally stripped are the most frequent causes of GCs being redistributed to the halo \citep{Kravtsov2005,Kruijssen2012,Rieder2013}. Using the E-MOSAICS suite of cosmological simulations, \cite{Kruijssen2015} and more recently \cite{Keller2020} have both shown that mergers play a vital role in the survival of GCs since they eject them to the outskirts where the interstellar medium is less dense and they are less prone to experience tidal shocks.  

It follows then that the connection between lower values of core radius for the GCS and galaxies that show substructures that imply major mergers not long in the past is a physical one. Large accretion events directly affect the radial distribution of GCs, not so much through the creation of young GCs but through the flattening of the inner region due to redistribution. In Table\,\ref{tab:gmm}, we show that the red fraction increases as we move away from the central region. In the current paradigm of GC formation, blue GCs are expected to dominate in the outer region since a large fraction of them comes from accretion of satellite galaxies, while most red GCs are thought to have formed in-situ. This difference can be interpreted as a sign of redistribution mechanisms at play.

\subsection{The bridge?}

The possibility of NGC\,3640 and NGC\,3641 interacting is not \rvw{thoroughly} discussed in previous work, mainly due to the observations focusing on the central regions NGC\,3640. In terms of integrated light, we do not detect a significant overdensity that can be interpreted as a stream between the two, but it could potentially be fainter than our magnitude limit. Since GCs have been shown to trace stellar streams \citep[e.g][]{Blom2014,Napolitano2022}, it is reasonable to interpret the overdensity detected between the two galaxies as a hint that there might be a stream between them. Previous works in other galaxies have also found potential bridges made of GCs as tracers of interactions, such as \cite{Bassino2006} connecting NGC\,1387 and NGC\,1399, and \cite{Wehner2008} with NGC\,3311 and NGC\,3309. We see that the colours of the GCs spatially situated between both galaxies are mainly intermediate, closer to the peak of the population associated with NGC\,3641, perhaps due to being members of its GCS that are being accreted by the more massive galaxy. As is originally mentioned in P88, the compactness of NGC\,3641 \rvw{might imply} a high angular momentum interaction. Although they refrain from interpreting anything more conclusive, they point out that the dust lane in NGC\,3640 is oriented in the direction of its neighbour, which is encouraging. 

Solid evidence of the interaction can only be obtained through spectroscopy, which would allow us to confirm membership of the GCs as well as to analyze their kinematics, which might confirm the connection between the two galaxies.

\section{Conclusions}

We perform a wide-field photometric study of the GCS of NGC\,3640 and NGC\,3641, using data obtained from GMOS/Gemini. We attribute GCs to the GCS of each galaxy based on projected distance, and analyze each of them separately. NGC\,3640 presents a bimodal GCS, with blue GCs that extend towards the outer regions of the field, and red GCs concentrated towards the centre. The luminosity function allows us to estimate a total population of $\approx 350$\,GCs. In NGC\,3641 we find a unimodal distribution, with GCs in an intermediate colour range in relation to the subpopulations in NGC\,3640. 

Finally, we identify a potential connection between the spatial distribution of the GCs and the shell-like structures around NGC\,3640, which hints at a major merger having disrupted the GCS. We also indicate the potential presence of a bridge formed by red GCs that connect both galaxies, indicating they may be currently interacting. Further spectroscopic studies are necessary to solidify these claims.

\section*{Acknowledgements}
We thank the referee for their careful reading of the paper, and for their valuable comments and suggestions. This research was supported in part by Perimeter Institute for Theoretical Physics.  Research at Perimeter Institute is supported by the Government of Canada through the  Department of Innovation, Science and Economic Development and by the Province of  Ontario through the Ministry of Research and Innovation. This work was funded with grants from Consejo Nacional de Investigaciones Cient\'ificas y T\'ecnicas de la Rep\'ublica Argentina, Agencia Nacional de Promoci\'on Cient\'fica y Tecnol\'ogica, and Universidad Nacional de La Plata (Argentina). This work made use of AstroPy \citep{astropy1,astropy2,astropy3}, MatPlotLib \citep{matplotlib}, NumPy \citep{numpy} Pandas \citep{pandas}, PhotUtils \citep{photutils}, SciKit \citep{scikit-learn} and SciPy \citep{scipy}.

\bibliographystyle{mnras}
\bibliography{biblio}



\end{document}